\newif\ifShowKeys
\ShowKeystrue
\ShowKeysfalse

\documentclass[letterpaper,10pt]{article}

\pdfoutput=1
\usepackage[no-natbib-sort]{my-jheppub}

\ifShowKeys \usepackage[notcite]{showkeys} \fi

\usepackage{amsmath, amssymb}

\usepackage{amsthm}

\usepackage{bm,environ,mathrsfs,array,arydshln,booktabs,float,slashed,hyperref}
\usepackage[mathcal]{euscript}
\usepackage{tensor} 	

\usepackage{wick}

\usepackage[nodayofweek]{date time}

\usepackage{graphicx,epsfig}
\usepackage{epic}
\usepackage{tikz} 
\usetikzlibrary{decorations.pathreplacing,decorations.markings,snakes}

\tikzset{middlearrow/.style={
        decoration={markings,
            mark= at position 0.5 with {\arrow{#1}} ,
        },
        postaction={decorate}
    }
}

\usepackage{floatflt}

\usepackage{aurical}
\usepackage[T1]{fontenc}

\usepackage{framed}
\definecolor{shadecolor}{RGB}{255, 230, 204}

\allowdisplaybreaks

\newcommand{\be}{\begin{equation}}
\newcommand{\ee}{\end{equation}}
\newcommand{\mc}{\mathcal }
\newcommand{\mb}{\mathbb }

\newcommand{\la}{\label}

\def \bz {\mathsf{z}}
\def \vp {\varphi}

\newcommand{\de}{\Delta}

\makeatletter
\DeclareFontFamily{OMX}{MnSymbolE}{}
\DeclareSymbolFont{MnLargeSymbols}{OMX}{MnSymbolE}{m}{n}
\SetSymbolFont{MnLargeSymbols}{bold}{OMX}{MnSymbolE}{b}{n}
\DeclareFontShape{OMX}{MnSymbolE}{m}{n}{
    <-6>  MnSymbolE5
   <6-7>  MnSymbolE6
   <7-8>  MnSymbolE7
   <8-9>  MnSymbolE8
   <9-10> MnSymbolE9
  <10-12> MnSymbolE10
  <12->   MnSymbolE12
}{}
\DeclareFontShape{OMX}{MnSymbolE}{b}{n}{
    <-6>  MnSymbolE-Bold5
   <6-7>  MnSymbolE-Bold6
   <7-8>  MnSymbolE-Bold7
   <8-9>  MnSymbolE-Bold8
   <9-10> MnSymbolE-Bold9
  <10-12> MnSymbolE-Bold10
  <12->   MnSymbolE-Bold12
}{}

\let\llangle\@undefined
\let\rrangle\@undefined
\DeclareMathDelimiter{\llangle}{\mathopen}%
                     {MnLargeSymbols}{'164}{MnLargeSymbols}{'164}
\DeclareMathDelimiter{\rrangle}{\mathclose}%
                     {MnLargeSymbols}{'171}{MnLargeSymbols}{'171}
\makeatother

\title{Toda theory in AdS$_{2}$ and $\mc WA_{n}$-algebra structure of  boundary correlators}

\author[a]{Matteo Beccaria and Giulio Landolfi} 

\abstract{We consider the conformal $A_{n}$  Toda theory in AdS$_{2}$.
Due to the bulk full Virasoro symmetry, this system
provides an instance of a non-gravitational AdS$_{2}$/CFT$_{1}$ correspondence where the 1d boundary theory enjoys
 enhanced ``$\frac{1}{2}$-Virasoro'' symmetry. General boundary correlators are expected to be captured
by the restriction of chiral correlators in a suitable $\mc WA_{n}$ Virasoro extension. 
At next-to-leading order in weak coupling expansion
they have been conjectured to match the subleading terms in the large central charge expansion 
of the dual $\mc WA_{n}$
correlators. 
We explicitly test this conjecture on the boundary four point functions
of the Toda  scalar fields dual to $\mc WA_{n}$ generators with  next-to-minimal spin 3 and 4. 
Our analysis is valid in the generic rank case and extends previous results  for 
specific rank-2 Toda theories.  
On the AdS side, the extension is straightforward and requires the computation
of a finite set of tree Witten diagrams. This is due to simple rank dependence and selection rules of cubic and quartic couplings.
On the boundary, we exploit crossing symmetry  
and specific meromorphic properties of the $\mc W$-algebra correlators at large central charge. 
We present the required 4-point functions in closed form for any rank and  verify the bulk-boundary 
correspondence in full details.
}

\affiliation[a]{Dipartimento di Matematica e Fisica Ennio De Giorgi,\\
Universit\`a del Salento \& INFN, Via Arnesano, 73100 Lecce, 
Italy} 
\emailAdd{matteo.beccaria@le.infn.it} 
\emailAdd{giulio.landolfi@le.infn.it}

\begin{document}


\maketitle


\section{Introduction}

Recently, the old subject of quantum field theories in  rigid AdS background
\cite{Callan:1989em}
turned out to be conveniently seen as a non-gravitational instance of AdS/CFT, suggesting new ideas and methods.
For example, flat space scattering amplitudes
of a massive theory may be obtained at large curvature radius  by studying 
the large scaling dimension regime of the boundary conformal correlators  
\cite{Paulos:2016fap,Paulos:2016but,Carmi:2018qzm}.
The specific   AdS$_{2}$ case  
attracted much interest from the very beginning
\cite{DHoker:1983zwg,DHoker:1983msr,Inami:1985di} and from the AdS/CFT perspective it has very 
special features, like the conjectured  duality  between a gravitational theory in AdS$_2$  and a 
chiral  half of a 2d CFT  \cite{Strominger:1998yg}. 
\footnote{
Recently, the rigid AdS$_{2}$ background
 played a key role in the analysis of correlators
of $\mathcal N=4$ SYM local operators inserted  on a  straight or circular Wilson line 
\cite{Drukker:2000ep,Alday:2007he,Polyakov:2000ti,Polchinski:2011im, 
Drukker:2006xg,Giombi:2017cqn,Beccaria:2018ocq,Beccaria:2019dws}. 
At strong coupling, the AdS$_{5}\times S^5$ string action is expanded 
near the   minimal surface associated with the 
1d defect and leads to a 2d field theory action in  AdS$_2$ background 
\cite{Giombi:2017cqn,Beccaria:2019dws}. 
}
In the gravitational 
context,  2d diffeomorphisms are a gauge symmetry and Virasoro symmetry 
may appear as an   asymptotic symmetry whose boundary manifestation 
are  1d time reparametrizations \cite{Hotta:1998iq,Cadoni:1999ja,NavarroSalas:1999up}, possibly 
spontaneously broken to $SL(2,\mathbb R)$ 
\cite{Almheiri:2014cka,Maldacena:2016upp,Engelsoy:2016xyb}.
 Instead, for a rigid AdS$_{2}$ background, the natural counterpart of this setup is to consider 
 a theory  that is locally conformal in the bulk and to explore the occurrence of 
enhanced boundary conformal symmetry. 
 As a first step in this direction, the analysis in \cite{Beccaria:2019stp} examined the case of  
Liouville theory \cite{Polyakov:1981rd,Teschner:2001rv,Nakayama:2004vk} with curved space action 
\be
\la{1.1}
\mc S = \frac{1}{4\pi}\int d^{2}x\,\sqrt{g}\, \big(\partial^{a}\varphi\partial_{a}\varphi
+\mu\,e^{2\,\beta\,\varphi}+Q\,R\,\varphi\big),\qquad\qquad  Q=\beta+\beta^{-1} .
\ee
It is Weyl-covariant on a fixed curved 2d background  with the central charge $c=1+6\,Q^{2}$.
In particular, on Euclidean AdS$_{2}$   background  with metric $ds^{2} = \frac{1}{\bz^{2}}(dt^{2}+d\bz^{2})$
 the Liouville  field $\vp$  can be expanded near its constant vacuum  expectation value and its fluctuations have classical 
 mass $m^2= 2$.  Bulk properties of Liouville theory on AdS$_{2}$ have been discussed previously 
 \cite{DHoker:1983msr,
 Zamolodchikov:2001ah,Menotti:2004uq}, while the recent  study in \cite{Beccaria:2019stp} focused
 on boundary correlators. With Dirichlet  boundary condition on the AdS boundary  $\bz=0$  the field $\vp$ has 
  asymptotics  $\varphi(t, \bz)\big|_{\bz\to 0} =\bz^2 \Phi (t) + ...$, and is dual to the 1d 
 CFT operator $\Phi(t)$ with  conformal  dimension $\Delta=2$ obeying the AdS$_{2}$ relation $m^{2}=\Delta(\Delta-1)$.
 The associated boundary correlators are defined as usual by 
\be 
\la{1.2}
\llangle \Phi(t_{1})\cdots \Phi(t_{n})\rrangle
    \equiv \lim_{\bz_1,...,\bz_n\to 0} \bz^{-2}_1 \cdots \bz^{-2}_n
    \, \langle\varphi(t_{1}, \bz_{1})\cdots\varphi(t_{n}, \bz_{n})\rangle,
    \ee
and can be computed perturbatively in the weakly coupled Toda theory 
by expanding in Witten diagrams. Since we start from a 2d conformal theory
in AdS$_{2}$, we can expect a correspondence between 
the boundary correlators and standard two-dimensional Virasoro correlators with the same
central charge. Indeed, as noticed in \cite{Ouyang:2019xdd}, this is true at tree level, {\em i.e.} 
$\beta\ll 1$ or $c\gg 1$. The 2-, 3-  and 4-point  boundary correlators (\ref{1.2}) match the correlators of the  
holomorphic stress tensor $T(z)$  according to 
  \be
\la{1.3}
\llangle \Phi(t_{1})\cdots \Phi(t_{n})\rrangle = \kappa^{n}\, \langle T(z_{1})\cdots T(z_{n})\rangle\Big|_{z_{i}\to t_{i}} \ ,
\ee
where $\kappa=\kappa(\beta) $ is a  proportionality coefficient appearing in the 
formal  identification $\Phi(t) \to  \kappa\, T(t)$ upon restriction of the 2d chiral stress tensor 
to the real  boundary $z_i=t_i + i y_i \to t_i$. \footnote{The identification can be explained at semiclassical level
by identifying $\Phi$ as the surviving piece in the boundary limit of the Toda stress tensor. This  simple
reasoning requires quantum refinements as discussed for the Liouville theory in \cite{Beccaria:2019stp}.}
In \cite{Beccaria:2019stp}, the relation 
(\ref{1.3}) has been tested beyond the leading tree level 
approximation by computing  the one-loop  corrections to  various correlators
$\llangle \Phi(t_{1})\cdots \Phi(t_{n})\rrangle$.
One of the outcomes of the analysis is the following proposal for the all-order expression of the intertwining coefficient $\kappa(\beta)$ 
\be
\la{1.4}
\kappa(\beta)  = -\frac{4Q}{c} = - \frac{ 4\,\beta  (1 + \beta^2)}{(3+2\,\beta^{2})(2+3\,\beta^{2})} = 
-\frac{2}{3}\,\beta+\frac{7}{9}\,	\beta^{3}+\cdots.
\ee
 A natural generalization 
  of the  Liouville correspondence (\ref{1.3}) consists in its extension  to 
  conformal Toda theories  of non-affine type   \cite{Gervais:1983am,Mansfield:1982sq,Braaten:1983pz}
on  the   AdS$_{2}$   background.
In the $A_{n}$ case,  expanding near the minimum of the Toda potential, 
one finds  $n$ scalar fields $\varphi_{\de}$ with masses $m^{2}=\de(\de-1)$   corresponding to 
  $\de=2, \dots, n+1$  \cite{Ouyang:2019xdd}. 
  The expected generalization of the duality relation  
  (\ref{1.3}) reads then 
\be
\la{1.5}
\llangle \Phi_{\de_{1}}(t_{1})\cdots \Phi_{\de_{n}}(t_{n})\rrangle =\big( \prod_{i=1}^{n}\kappa_{\de_{i}}\big) \ 
\langle Q_{\de_{1}}(z_{1})\cdots Q_{\de_{n}}(z_{n})\rangle\Big|_{z_{i}\to t_{i}} \ ,
\ee
where  $\llangle \Phi_{\de_{1}}(t_{1})\cdots \Phi_{\de_{n}}(t_{n})\rrangle= 
\lim_{\bz_1,...,\bz_n\to 0} \bz^{-\de_{1}}_1 \cdots \bz^{-\de_{n}}_n
    \, \langle\varphi_{\de_{1}}(t_{1}, \bz_{1})\cdots\varphi_{\de_{n}}(t_{n}, \bz_{n})\rangle$,
$Q_\de= \{Q_{2}\equiv T, Q_{3}, \dots, Q_{n+1}\}$ are the generators of the chiral $\mc W_{n+1}$ algebra  
replacing and extending  the Virasoro symmetry
 and  with  the same central charge of the Toda theory. The coefficients $\kappa_{\de_{i}}$
are functions of the Toda coupling entering the correspondence $\Phi_{\de}\to \kappa_{\de}Q_{\de}$.
The relation (\ref{1.5}) was  noticed  at  tree level in \cite{Ouyang:2019xdd} 
in a few sample 4-point functions
in the Toda theories  associated to  some  rank-2 algebras
 with two scalar fields (one dual to the
stress  tensor $T$ and the other dual to a higher spin chiral field $Q_{s}$). 

In this paper, we discuss the relation (\ref{1.5})
for the four point functions 
involving the two  fields with next-to-minimal higher 'spin' $\de=3,4$ in the general $A_{n}$ Toda theory.
This analysis aims to exclude possible low-rank accidental good properties. Despite being a leading order
analysis, not involving loops in AdS, \footnote{
It is natural to expect (\ref{1.5})
to hold also at the  quantum level  as 
should be possible to check   by the methods  used  in \cite{Beccaria:2019stp}, {\em cf.}  \cite{BHT}.}
the AdS/CFT matching of the full  dependence on the rank $n$ proves to be a quite stringent constraint. 

Technically, the $A_{n}$ case is feasible and rather straightforward on the AdS side due to some peculiar regularities of 
the cubic and quartic couplings with respect to the rank  $n$. At leading non-trivial order, 
selection rules reduce the calculation to a finite sum of
Witten diagrams that can be exactly computed. On the CFT side, the task is in principle harder and amounts to 
the calculation of 4-point correlators of spin-3 and spin-4 generators of the Casimir W-algebra 
$\mc W_{n+1}$ \cite{Bouwknegt:1992wg}. The structure of such algebras depends non-trivially on the rank $n$
and the fusion structure constants are  functions of $n$ and the central charge that are not known 
in general. Nevertheless, we shall be only interested in the leading and sub-leading correlators at large central charge.
Known results about semiclassical Virasoro blocks together with a careful use of crossing symmetry and the meromorphic
properties of the correlators will allow a simple determination of the desired correlators for generic rank. \footnote{
Clearly, at given rank, one can use the explicit fusion algebra of $\mc WA_{n}$ or, what is the same, compute the four-point
functions using free-field representations, {\em cf.} \cite{Fateev:1987zh}. However, as we pointed out, 
our interest is focused on the 
generic rank subleading
corrections to the large charge limit and a more direct approach will be more convenient to this aim.
}
Our analysis confirms the validity of (\ref{1.5}) for the  considered 4-point functions  
in the $A_{n}$ Toda theory, at least at classical level. This lends 
further support to that relation and, in principle, allows to test higher loop AdS calculations 
by $\mc W$-algebraic methods.

The structure of the paper is as follows. In  Section \ref{sec:toda} we illustrate the tools needed to compute 
tree level boundary correlators in the $A_{n}$ Toda theory on AdS$_{2}$ and present explicit results
for the 4-point functions of scalars with $\Delta=3,4$. Section \ref{sec:w-corr} presents the associated results
for the dual CFT fields in the $\mc W_{n}$ Virasoro extension. The correlators of generators with spin 3, 4
are computed for generic rank at subleading order in the large central charge expansion. Finally, 
Section \ref{sec:final} concludes the analysis by checking agreement with the universal relation (\ref{1.5}).
Some technical tools are briefly collected in two appendices.

\section{Tree-level 4-point functions in $A_{n}$ Toda theory in AdS$_{2}$}
\la{sec:toda}

We shall consider the $A_{n}$ Toda theory in AdS$_{2}$ with classical action,
see for instance  \cite{Fateev:2007ab},
\be
\la{2.1}
\mc S_{n} = \int d^{2}x\,\sqrt{g}\,\bigg[\tfrac{1}{2}\partial^{\mu}\bm\phi\cdot\partial_{\mu}\bm\phi+
V_{n}(\bm \phi)\bigg],\qquad 
V_{n}(\bm \phi) = \frac{1}{\beta^{2}}\sum_{i=1}^{n}q_{i}\,e^{\beta\,\bm\alpha_{i}\cdot\bm\phi}
+\frac{1}{\beta}\,R\,\bm \rho^{\vee}\cdot \bm \phi,
\ee
where $\beta$ is a coupling and $V_{n}$ will be referred to as the {\em potential}. 
In the special case of $A_{1}$ the action reduces to the Liouville action, {\em cf.} (\ref{1.1}).
The field $\bm\phi$ is an $n$-component multiplet of scalar fields, $\bm\alpha_{i}$ are the simple roots of the Lie algebra
$A_{n}$ and the Weyl vector $\bm\rho^{\vee}$ satisfies $\bm\alpha_{i}
\cdot\bm\rho^{\vee}=1$ for all $i=1,\dots, n$. The numbers $q_{i}$ are taken to be the unique solution to the condition
$\sum_{i=1}^{n}q_{i}\ \bm\alpha_{i}\cdot \bm\alpha_{j} = 2$, for  $j=1, \dots, n$.
\footnote{The numbers $q_{i}$ are not restricted to be  
integer since a shift of the scalar fields is understood and such that linear terms are removed from the action.}
The action (\ref{2.1}) is  Weyl invariant \footnote{This requires a quantum shift $\frac{1}{\beta}\to 
\frac{1}{\beta}+\beta$ in the last term of the potential in (\ref{2.1}). This correction will have no effects 
at the perturbative order of the calculations in this paper. Instead, it is an important ingredient in the 
loop corrections discussed in \cite{Beccaria:2019stp}.}
 and flat space integrability carries over to any 
conformally flat background. In particular we are interested in AdS$_{2}$ with unit radius and Poincar\'e coordinates 
\be
ds^{2} = \tfrac{1}{z^{2}}\,(dz^{2}+dt^{2}),	\qquad (t,z)\in\mathbb{R}\times \mathbb{R}^{+}.
\ee
The kinetic part of the action (including  mass terms) is diagonalized by going to a basis of normalized eigenvectors
of the matrix $\mc A_{ij} = q_{k} (A^{1/2})_{ki}\,(A^{1/2})_{kj}$, 
where $A$ is the (symmetric) Cartan matrix
of the $A_{n}$ algebra
\be
\la{2.3}
A_{ij} = \frac{2\,\bm\alpha_{i}\cdot \bm\alpha_{j}}{|\bm\alpha_{i}|^{2}} = \tfrac{1}{2}\,\bm\alpha_{i}\cdot \bm \alpha_{j},
\qquad 
A = {\footnotesize \begin{pmatrix} 
2 & -1 & 0 & \cdots & 0 \\
-1 & 2 & -1 & \cdots & 0 \\
& \cdots & \cdots \\
0 & \cdots & -1  & 2 & -1 \\
0 & \cdots & 0 & -1  & 2
\end{pmatrix}}.
\ee
The mass matrix $\mc A_{ij}$ can be put in a simple tridiagonal form. To this aim, one
writes the $A_{n}$ simple roots in the form 
$\bm\alpha_{1} = \sqrt{2}\,\bm e_{1}$,  $\bm \alpha_{2} = \beta_{21}\,\bm e_{1}+\beta_{22}\,\bm e_{2}$, $\dots$,  
$\bm \alpha_{p} = \sum_{q=1}^{p}\beta_{pq}\bm e_{q}$,
where $\bm e_{1}, \dots, \bm e_{n}$ are orthonormal vectors in $\mb R^{n}$. 
This gives the only non-zero elements
\be
\mc A_{p, p} = \frac{n+1}{2}+n\,p-p^{2},\qquad 
\mc A_{p, p+1}^{2} = \mc A_{p+1, p}^{2} = \frac{1}{4}\,p\,(p+2)\,(p-n)^{2}.\notag 
\ee
In terms of the rotated fields $\bm\phi\to \bm\phi'$, the Lagrangian reads 
\be
\mc L = \tfrac{1}{2}\,\partial^{\mu}\bm\phi'\cdot\partial_{\mu}\bm\phi'+\tfrac{1}{2}\,\sum_{i=1}^{n}\,
m_{i}^{2}\,(\phi_{i}')^{2}+V_{n}(\bm\phi'),
\ee
where the masses can be written $m^{2}_{i} = \de_{i}\,(\de_{i}-1)$ with the simple pattern
\be
\{\de_{i}\} = 2, 3, 4, \dots, n+1.
\ee
According to the AdS/CFT dictionary, $\de_{i}$ is the conformal dimensions of the dual boundary fields. 
Since they will turn out to be chiral fields, we shall often refer to $\de_{i}$ as the {\rm spin} quantum number.
The non-polynomial potentials $V_{n}(\bm\phi')$ can be expanded in powers of $\beta$ producing cubic, quartic and higher order
couplings. 
It is convenient to relabel the fields by using their would-be conformal dimension
$\phi_{i}'\to \varphi_{\de_{i}}$. Besides, for the following discussion, 
it will be convenient to rescale $\beta\to \sqrt{\frac{n(n+1)(n+2)}{6}}\,\beta$, depending on $n$, 
in order to have the same $\varphi_{2}^{3}$ coupling for all $n$. With these 
conventions, in the first three cases $n=1,2,3$,  the explicit potentials read
\begin{align}
\la{2.6}
V_{1}(\varphi_{2}) &= \varphi_{2}^2+\tfrac{2}{3}\,\beta\,\varphi_2^3+\tfrac{1}{3} \beta ^2 \varphi_{2}^4+\cdots\,,  \\
V_{2}(\varphi_{2}, \varphi_{3}) &=  \varphi _2^2+3 \varphi _3^2+\beta  (\tfrac{2}{3} \varphi _2^3+6 \varphi _3^2 
\varphi _2) +\beta ^2 (\tfrac{1}{3} \varphi _2^4+6 \varphi _3^2 \varphi _2^2+3 \varphi_3^4)+\cdots\,, \notag \\
V_{3}(\varphi_{2}, \varphi_{3}, \varphi_{4}) &=
\varphi _2^2+3 \varphi _3^2+6 \varphi _4^2+\beta 
 (\tfrac{2}{3} \varphi _2^3+6 
\varphi _3^2 \varphi _2+12 \varphi _4^2 \varphi _2-4 \varphi _4^3+12 \varphi _3^2 \varphi_4)\notag \\
& +\beta ^2 (\tfrac{1}{3}\,\varphi _2^4+6 \varphi _3^2 \varphi _2^2+12 \varphi _4^2 
\varphi _2^2-8 \varphi _4^3 \varphi _2+24 \varphi _3^2 \varphi _4 \varphi _2+5 \varphi_3^4
+14 \varphi _4^4+24 \varphi _3^2 \varphi _4^2)
+\cdots\, .\notag
\end{align}

\bigskip\noindent
We want to discuss the 4-point function of the scalars dual to the first two 
higher spin fields,  those with spin 3 and 4. The non-zero cases are the two diagonal
 and the one mixed  4-point functions
 \be
\la{2.7}
\llangle\varphi_{3}\varphi_{3}\varphi_{3}\varphi_{3}\rrangle,\qquad
\llangle\varphi_{3}\varphi_{3}\varphi_{4}\varphi_{4}\rrangle,\qquad
\llangle\varphi_{4}\varphi_{4}\varphi_{4}\varphi_{4}\rrangle.
\ee
To compute them in the generic $A_{n}$ Toda theory, we need some of the cubic and quartic couplings at 
generic $n$.  Not surprisingly, the non zero couplings we shall need are only a finite set. Apart from the 
$3333$, $3344$, and $4444$ contact Witten diagrams, we can have an exchange in one of three kinematical channels
mediated by two cubic interactions. The $33s$ couplings are non-zero only for $s=2,4$ while
the $44s$ couplings are non-zero only for $s=2,4,6$. Finally, for the $3344$ four-point function 
we also need the non-zero couplings $34s$. Apart from $s=3$, there is only $s=5$.
A systematic analysis of the potentials at increasing rank shows that we can write  
\begin{align}
\la{2.8}
V_{n}(\bm\varphi) &= \cdots+\beta\,(\tfrac{2}{3}\,\varphi_{2}^{3}+6\,\varphi_{3}^{2}\,\varphi_{2}
+12\,\varphi_{4}^{2}\,\varphi_{2}
+c_{334}\,\varphi_{3}^{2}\,\varphi_{4}
+c_{444}\,\varphi_{4}^{3}
+c_{446}\,\varphi_{4}^{2}\,\varphi_{6}
+c_{345}\,\varphi_{3}\,\varphi_{4}\,\varphi_{5}
)\notag \\
& +
\beta^{2}\,(c_{3333}\,\varphi_{3}^{4}+c_{3344}\,\varphi_{3}^{2}\,\varphi_{4}^{2}+
c_{4444}\,\varphi_{4}^{4})
+\cdots,
\end{align}
where the couplings are expressed by the following remarkably simple rational expressions 
\footnote{The formulas for couplings involving the field $\varphi_{k}$ 
are valid if  $n\ge k-1$, otherwise that field is simply absent.}
\begin{align}
\la{2.9}
(c_{334})^{2}  &= \frac{1728}{7}\,\frac{(n+4)(n-2)}{(n+3)(n-1)}, &
c_{3333} &= \frac{15}{7}\,\frac{3n^{2}+6n-17}{(n+3)(n-1)}, \notag \\
(c_{444})^{2}  &= \frac{448 (n^2+2 n-18)^2}{3 (n-2) (n-1) (n+3) (n+4)} , &
(c_{446})^{2} &= \frac{30000 (n-4) (n-3) (n+5) (n+6)}{11 (n-2) (n-1) (n+3) (n+4)} , \notag \\
c_{3344}  &= \frac{12 (7 n^2+14 n-81)}{(n-1) (n+3)}, &
c_{4444} &= \frac{56 (9 n^4+36 n^3-238 n^2-548 n+2316)}{11 (n-2) (n-1) (n+3) (n+4)},\notag \\
(c_{345})^{2} &= \frac{24000 (n-3) (n+5)}{7 (n-1) (n+3)}.
\end{align}
Notice that in the case of affine Toda theories, the coupling structure is somewhat simpler and there are 
known selection rules for the cubic couplings, as well as (recursion) relations for their values 
\cite{Christe:1989ah,Christe:1989my,Braden:1989bu,Braden:1989bg,Gabai:2018tmm}. Expressions (\ref{2.9}) 
are valid for the finite Lie algebra Toda theories and are novel necessary ingredients.

\medskip
\noindent
To compute 4-point functions like (\ref{2.7}) at leading order, we need the 
bulk-to-bulk propagator for a scalar field in AdS$_{2}$ with mass term $m^{2}=\de(\de-1)$. It is 
\begin{align}
\la{2.10}
& G_{\de}(t_{1}, \bz_{1}; t_{2}, \bz_{2}) = \mc C_{\de}\,u^{-\de}\,_{2}F_{1}(\de, \de, 2\de; -\tfrac{4}{u}),\quad 
\mc C_{\de} = \tfrac{\Gamma(\de)}{2\,\sqrt\pi\,\Gamma(\de+\frac{1}{2})},
\end{align}
where the chordal distance is $u = \frac{(t_{1}-t_{2})^{2}+(\bz_{1}-\bz_{2})^{2}}{\bz_{1}\,\bz_{2}}$. 
The bulk-to-boundary propagator is, after suitable field rescaling,  
\be
K_{\de}(t_{0}; t, \bz) = \mc C_{\de}\left[\frac{\bz}{\bz^{2}+(t-t_{0})^{2}}\right]^{\de},
\ee
and is associated with fields whose 2-point function has coefficient $\mc C_{\de}$. Since we shall be interested in 
fields with unit normalized 2-point function, one factor $\mc C_{\de_{i}}^{-1/2}$ will be attached to any boundary-to-bulk
line. Some technical tools for the evaluation of AdS integrals are collected in App. \ref{app-Dfun}.

\subsection{Four point functions involving $\de=3,4$ fields in the $A_{n}$ Toda theory}

We now compute the 4-point functions (\ref{2.7}) by evaluating the associated Witten diagrams and using
the general couplings in (\ref{2.9}). The leading order will be given by two boundary-to-boundary propagators
and is a disconnected contribution independent on $\beta$.  This part is almost trivial and will be discussed in the end. 
Instead, we focus on the non-trivial connected contribution. At leading order, this starts at quadratic order $\mc O(\beta^{2})$.

\subsubsection{$\de=3$ boundary correlator $\llangle \varphi_{3}\varphi_{3}\varphi_{3}\varphi_{3}\rrangle$}

\begin{figure}[ht]
\centering
\begin{tikzpicture}[line width=1 pt, scale=0.5, rotate=0]
\coordinate (A1) at (140:2);
\coordinate (A2) at (-140:2);
\coordinate (A3) at (40:2);
\coordinate (A4) at (-40:2);
\coordinate (M1) at (-1,0);    \coordinate (M2) at (1,0);

\draw[densely dashed,line width=0.5pt] (0,0) circle (2);
\draw (A1)--(M1)--(A2); \draw (A3)--(M2)--(A4); \draw (M1)--(M2);

\draw[fill=black] (A1) circle (0.12); \node[left] at (A1) {$3$};
\draw[fill=black] (A2) circle (0.12); \node[left] at (A2) {$3$};
\draw[fill=black] (A3) circle (0.12); \node[right] at (A3) {$3$};
\draw[fill=black] (A4) circle (0.12); \node[right] at (A4) {$3$};


\node[above] at (0,0) {$2, 4$};
\node at (3,0) {$+$};
\end{tikzpicture}
\begin{tikzpicture}[line width=1 pt, scale=0.5, rotate=0]
\coordinate (A1) at (140:2);
\coordinate (A2) at (-140:2);
\coordinate (A3) at (40:2);
\coordinate (A4) at (-40:2);
\coordinate (M1) at (0,0.8);    \coordinate (M2) at (0,-0.8);

\draw[densely dashed,line width=0.5pt] (0,0) circle (2);
\draw (A1)--(M1)--(A3); \draw (A2)--(M2)--(A4); \draw (M1)--(M2);

\draw[fill=black] (A1) circle (0.12); \node[left] at (A1) {$3$};
\draw[fill=black] (A2) circle (0.12); \node[left] at (A2) {$3$};
\draw[fill=black] (A3) circle (0.12); \node[right] at (A3) {$3$};
\draw[fill=black] (A4) circle (0.12); \node[right] at (A4) {$3$};


\node[left] at (0,0) {$2, 4$};
\node at (4,0) {+ \scriptsize u-channel};
\end{tikzpicture}
\begin{tikzpicture}[line width=1 pt, scale=0.5, rotate=0]
\coordinate (A1) at (140:2);
\coordinate (A2) at (-140:2);
\coordinate (A3) at (40:2);
\coordinate (A4) at (-40:2);
\coordinate (M1) at (0,0.8);    \coordinate (M2) at (0,-0.8);

\draw[densely dashed,line width=0.5pt] (0,0) circle (2);
\draw (A1)--(0,0)--(A3); \draw (A2)--(0,0)--(A4); 
\draw[fill=black] (A1) circle (0.12); \node[left] at (A1) {$3$};
\draw[fill=black] (A2) circle (0.12); \node[left] at (A2) {$3$};
\draw[fill=black] (A3) circle (0.12); \node[right] at (A3) {$3$};
\draw[fill=black] (A4) circle (0.12); \node[right] at (A4) {$3$};


\end{tikzpicture}
\caption{Tree diagrams contributing $\llangle\varphi_{3}\varphi_{3}\varphi_{3}\varphi_{3}\rrangle$.
}\label{fig:3333}
\end{figure}
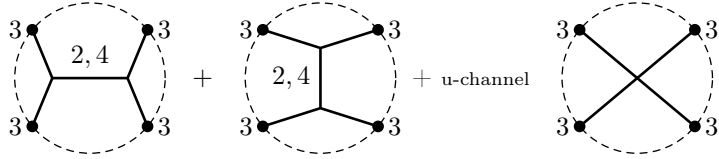

From the couplings in  (\ref{2.8}), we have the following  (Witten) Feynman rules for the relevant cubic vertices 
and quartic coupling 
\begin{center}
\begin{tikzpicture}[line width=1 pt, scale=0.6]
\draw (0:1)--(0,0);       \node[right] at (0:1) {2};
\draw (120:1)--(0,0);   \node[left] at (120:1) {3};
\draw (-120:1)--(0,0);   \node[left] at (-120:1) {3};
\node[right=0.3cm] at (0:1) {$= -12\,\beta,$};
\end{tikzpicture}
\hskip 1.5cm
\begin{tikzpicture}[line width=1 pt, scale=0.6]
\draw (0:1)--(0,0);       \node[right] at (0:1) {4};
\draw (120:1)--(0,0);   \node[left] at (120:1) {3};
\draw (-120:1)--(0,0);   \node[left] at (-120:1) {3};
\node[right=0.3cm] at (0:1) {$= -2\,\beta\,c_{334},$};
\end{tikzpicture}
\hskip 1.5cm
\begin{tikzpicture}[line width=1 pt, scale=0.6]
\draw (45:1)--(0,0);   \node[right] at (45:1) {3};
\draw (135:1)--(0,0); \node[left] at (135:1) {3};
\draw (225:1)--(0,0); \node[left] at (225:1) {3};
\draw (315:1)--(0,0);  \node[right] at (315:1) {3};
\node[right=0.2cm] at (0:1) {$= -4!\,\beta^{2}\,c_{3333},$};
\end{tikzpicture}
\end{center}
where the value of the couplings $c_{334}$ and $c_{3333}$ has been given in (\ref{2.9}). 
The four-point function is then given by the sum of diagrams in Fig.~(\ref{fig:3333}).
It can be computed in terms of the $\overline{D}$ functions defined in App.~(\ref{app-Dfun}). We find 
the connected contribution
\begin{align}
\mc C_{3}^{-2}\,\llangle & \varphi_{3}(t_{1})\cdots\varphi_{3}(t_{4})\rrangle_{\rm conn} = 
(-12\beta)^{2}(W^{s}_{3333; 2}+W^{t}_{3333; 2}+W^{u}_{3333; 2}) \notag \\
& +(-2\,\beta\,c_{334})^{2}\,(W^{s}_{3333; 4}+W^{t}_{3333; 4}+W^{u}_{3333; 4}) -4!\,
\beta^{2}\,c_{3333}\,D_{3333}\notag \\
&= \frac{15\,\pi\,\beta^{2}}{512}\,\bigg[
\frac{7 (c_{334}^2+36) 
\overline{D}_{2,2,3,3}}{t_{12}^2 t_{13}^4 
t_{24}^4 t_{34}^2}+\frac{7 (c_{334}^2+36) 
\overline{D}_{2,3,2,3}}{t_{13}^6 t_{24}^6}+\frac{7 (c_{334}^2+36) 
\overline{D}_{2,3,3,2}}{t_{13}^6 t_{24}^6}-\frac{756 c_{3333} 
\overline{D}_{3,3,3,3}}{t_{13}^6 t_{24}^6}\notag \\
& +\frac{180 
\overline{D}_{1,1,3,3}}{t_{12}^4 t_{13}^2 t_{24}^2 t_{34}^4}+\frac{180 
\overline{D}_{1,3,1,3}}{t_{13}^6 t_{24}^6}+\frac{180 
\overline{D}_{1,3,3,1}}{t_{13}^6 t_{24}^6}
\bigg].
\end{align}
Using the explicit values of the relevant $\overline{D}$ functions, {\em cf.} (\ref{B.4}), we may write 
\be
\mc C_{3}^{-2}\,
\llangle  \varphi_{3}(t_{1})\cdots\varphi_{3}(t_{4})\rrangle_{\rm conn} = \frac{\beta^{2}}{t_{12}^{6}\,t_{34}^{6}}\,
G_{3333}^{\rm AdS}(\chi),\qquad \chi =\frac{t_{12}t_{34}}{t_{13}t_{24}},
\ee
with 
\begin{align}
G_{3333}^{\rm AdS}(\chi) &= -(c_{334}^2-72 c_{3333}+216)\,\frac{3 \pi   \chi ^6 (\chi ^2-\chi +1) 
(2 \chi ^2-7 \chi +7)}{256 (1-\chi)^5}\,\log\chi\notag \\
&-\frac{3}{256} \pi  (c_{334}^2-72 c_{3333}+216) \chi  (\chi ^2-\chi +1) (2 \chi ^2+3 \chi +2)\,\log(1-\chi)\notag \\
& -\frac{3 \pi  \chi ^2}{1024 (1-\chi )^4} \bigg[c_{334}^2 (8 \chi ^6-24 \chi ^5+13 \chi ^4+14 
\chi ^3+13 \chi ^2-24 \chi +8)\notag \\
& -48\, c_{3333} (12 \chi ^6-36 \chi ^5+37 
\chi ^4-14 \chi ^3+37 \chi ^2-36 \chi +12)\notag \\
& -36 (2 \chi ^6-6 \chi ^5-3 
\chi ^4+16 \chi ^3-3 \chi ^2-6 \chi +2)\bigg].
\end{align}
The logarithmic terms $\sim\log\chi, \log(1-\chi)$
vanish using the explicit form of $c_{334}$ and $c_{3333}$, see (\ref{2.9}). 
This non-trivial fact will have an explanation in terms
of the boundary CFT and will be associated with the absence of anomalous dimensions of the various dual fields. 
The remaining expression for the 4-point function takes then the following compact form
\begin{align}
G_{3333}^{\rm AdS}(\chi) &= \frac{675\,\pi}{256\,(n-1)(n+3)}
\,\frac{\chi^{2}}{(1-\chi)^{4}}\,\bigg[
2 (n-1) (n+3)(1-3\chi-3\chi^{5}+\chi^{6})\notag \\
&  +(9 n^2+18 n-43) \chi ^2(1+\chi^{2}) -8 (n^2+2 
n-7) \chi ^3
\bigg],
\end{align}
that obeys the correct crossing relations
\be
G_{3333}^{\rm AdS}(\chi) = \tfrac{\chi^{6}}{(1-\chi)^{6}}\,G_{3333}^{\rm AdS}(1-\chi), \qquad 
G_{3333}^{\rm AdS}(\chi) = G_{3333}^{\rm AdS}(\tfrac{\chi}{\chi-1}).
\ee

\subsubsection{$\de=4$ boundary correlator $\llangle \varphi_{4}\varphi_{4}\varphi_{4}\varphi_{4}\rrangle$}

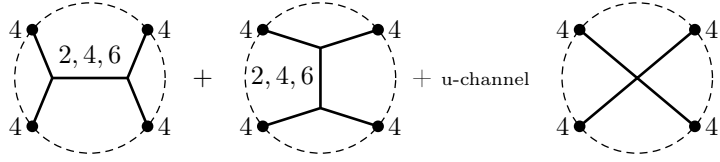
\begin{figure}[ht]
\centering
\begin{tikzpicture}[line width=1 pt, scale=0.5, rotate=0]
\coordinate (A1) at (140:2);
\coordinate (A2) at (-140:2);
\coordinate (A3) at (40:2);
\coordinate (A4) at (-40:2);
\coordinate (M1) at (-1,0);    \coordinate (M2) at (1,0);

\draw[densely dashed,line width=0.5pt] (0,0) circle (2);
\draw (A1)--(M1)--(A2); \draw (A3)--(M2)--(A4); \draw (M1)--(M2);

\draw[fill=black] (A1) circle (0.12); \node[left] at (A1) {$4$};
\draw[fill=black] (A2) circle (0.12); \node[left] at (A2) {$4$};
\draw[fill=black] (A3) circle (0.12); \node[right] at (A3) {$4$};
\draw[fill=black] (A4) circle (0.12); \node[right] at (A4) {$4$};


\node[above] at (0,0) {$2, 4, 6$};
\node at (3,0) {$+$};
\end{tikzpicture}
\begin{tikzpicture}[line width=1 pt, scale=0.5, rotate=0]
\coordinate (A1) at (140:2);
\coordinate (A2) at (-140:2);
\coordinate (A3) at (40:2);
\coordinate (A4) at (-40:2);
\coordinate (M1) at (0,0.8);    \coordinate (M2) at (0,-0.8);

\draw[densely dashed,line width=0.5pt] (0,0) circle (2);
\draw (A1)--(M1)--(A3); \draw (A2)--(M2)--(A4); \draw (M1)--(M2);

\draw[fill=black] (A1) circle (0.12); \node[left] at (A1) {$4$};
\draw[fill=black] (A2) circle (0.12); \node[left] at (A2) {$4$};
\draw[fill=black] (A3) circle (0.12); \node[right] at (A3) {$4$};
\draw[fill=black] (A4) circle (0.12); \node[right] at (A4) {$4$};


\node[left=-0.05cm] at (0,0) {$2, 4, 6$};
\node at (4,0) {+ \scriptsize u-channel};
\end{tikzpicture}
\begin{tikzpicture}[line width=1 pt, scale=0.5, rotate=0]
\coordinate (A1) at (140:2);
\coordinate (A2) at (-140:2);
\coordinate (A3) at (40:2);
\coordinate (A4) at (-40:2);
\coordinate (M1) at (0,0.8);    \coordinate (M2) at (0,-0.8);

\draw[densely dashed,line width=0.5pt] (0,0) circle (2);
\draw (A1)--(0,0)--(A3); \draw (A2)--(0,0)--(A4); 
\draw[fill=black] (A1) circle (0.12); \node[left] at (A1) {$4$};
\draw[fill=black] (A2) circle (0.12); \node[left] at (A2) {$4$};
\draw[fill=black] (A3) circle (0.12); \node[right] at (A3) {$4$};
\draw[fill=black] (A4) circle (0.12); \node[right] at (A4) {$4$};


\end{tikzpicture}
\caption{Tree diagrams contributing $\llangle\varphi_{4}\varphi_{4}\varphi_{4}\varphi_{4}\rrangle$.
}\label{fig:4444}
\end{figure}

In this case, again from (\ref{2.8}), we have the following cubic vertices 
and quartic coupling 
\begin{center}
\begin{tikzpicture}[line width=1 pt, scale=0.6]
\draw (0:1)--(0,0);       \node[right] at (0:1) {2};
\draw (120:1)--(0,0);   \node[left] at (120:1) {4};
\draw (-120:1)--(0,0);   \node[left] at (-120:1) {4};
\node[right=0.3cm] at (0:1) {$= -24\,\beta,$};
\end{tikzpicture}
\hskip 0.1cm
\begin{tikzpicture}[line width=1 pt, scale=0.6]
\draw (0:1)--(0,0);       \node[right] at (0:1) {4};
\draw (120:1)--(0,0);   \node[left] at (120:1) {4};
\draw (-120:1)--(0,0);   \node[left] at (-120:1) {4};
\node[right=0.3cm] at (0:1) {$= -3!\,\beta\,c_{444},$};
\end{tikzpicture}
\hskip 0.1cm
\begin{tikzpicture}[line width=1 pt, scale=0.6]
\draw (0:1)--(0,0);       \node[right] at (0:1) {6};
\draw (120:1)--(0,0);   \node[left] at (120:1) {4};
\draw (-120:1)--(0,0);   \node[left] at (-120:1) {4};
\node[right=0.3cm] at (0:1) {$= -2\,\beta\,c_{446},$};
\end{tikzpicture}
\hskip 0.1cm
\begin{tikzpicture}[line width=1 pt, scale=0.6]
\draw (45:1)--(0,0);   \node[right] at (45:1) {4};
\draw (135:1)--(0,0); \node[left] at (135:1) {4};
\draw (225:1)--(0,0); \node[left] at (225:1) {4};
\draw (315:1)--(0,0);  \node[right] at (315:1) {4};
\node[right=0.2cm] at (0:1) {$= -4!\,\beta^{2}\,c_{4444},$};
\end{tikzpicture}
\end{center}
where the couplings $c_{444}$, $c_{446}$, and $c_{4444}$ have been given in (\ref{2.9}). 
The four-point (connected) function is then given by the diagrams in Fig.~(\ref{fig:4444}) and reads 
\begin{align}
\mc C_{4}^{-2}\,\llangle & \varphi_{4}(t_{1})\cdots\varphi_{4}(t_{4})\rrangle_{\rm conn} = 
(-24\beta)^{2}(W^{s}_{4444; 2}+W^{t}_{4444; 2}+W^{u}_{4444; 2}) \notag \\
& +(-3!\,\beta\,c_{444})^{2}\,(W^{s}_{4444; 4}+W^{t}_{4444; 4}+W^{u}_{4444; 4}) \notag \\
& +(-2\,\beta\,c_{446})^{2}\,(W^{s}_{4444; 6}+W^{t}_{4444; 6}+W^{u}_{4444; 6}) 
-4!\,\beta^{2}\,c_{4444}\,D_{4444}\notag \\
&= \frac{\pi\,\beta^{2}}{6144}\,\bigg[
\frac{315 (9 c_{444}^2+224) \overline{D}_{2,2,4,4}}{t_{12}^4 t_{13}^4 
t_{24}^4 t_{34}^4}+\frac{315 (9 c_{444}^2+224) 
\overline{D}_{2,4,2,4}}{t_{13}^8 t_{24}^8}+\frac{315 (9 c_{444}^2+224) 
\overline{D}_{2,4,4,2}}{t_{13}^8 t_{24}^8}\notag \\
& +\frac{385 (9 
c_{444}^2+c_{446}^2+144) \overline{D}_{3,3,4,4}}{t_{12}^2 t_{13}^6 
t_{24}^6 t_{34}^2}+\frac{385 (9 c_{444}^2+c_{446}^2+144) 
\overline{D}_{3,4,3,4}}{t_{13}^8 t_{24}^8}+\frac{385 (9 
c_{444}^2+c_{446}^2+144) \overline{D}_{3,4,4,3}}{t_{13}^8 
t_{24}^8}\notag \\
& -\frac{60060 c_{4444} \overline{D}_{4,4,4,4}}{t_{13}^8 
t_{24}^8}+\frac{39200 \overline{D}_{1,1,4,4}}{t_{12}^6 t_{13}^2 
t_{24}^2 t_{34}^6}+\frac{39200 \overline{D}_{1,4,1,4}}{t_{13}^8 
t_{24}^8}+\frac{39200 \overline{D}_{1,4,4,1}}{t_{13}^8 t_{24}^8}
\bigg].
\end{align}
This may be written 
\be
\mc C_{4}^{-2}\,
\llangle  \varphi_{4}(t_{1})\cdots\varphi_{4}(t_{4})\rrangle_{\rm conn} = \frac{\beta^{2}}{t_{12}^{8}\,t_{34}^{8}}\,
G_{4444}^{\rm AdS}(\chi),
\ee
where 
\begin{align}
& G_{4444}^{\rm AdS}(\chi) = \frac{\pi\,\chi^{8}}{6144\,(1-\chi)^{7}}\,
\bigg[
-5121 c_{444}^2-110 c_{446}^2+30888 c_{4444}-350496\notag \\
& +3 (5121 
c_{444}^2+110 c_{446}^2-30888 c_{4444}+350496) \chi \notag \\
& +(-26019 
c_{444}^2-770 c_{446}^2+175032 c_{4444}-2034144) \chi ^2\notag \\
& +99 (267 
c_{444}^2+10 c_{446}^2-1976 c_{4444}+23392) \chi ^3-2 (8253 
c_{444}^2+350 c_{446}^2-64584 c_{4444}+772128) \chi ^4\notag \\
& +130 (45 
c_{444}^2+2 c_{446}^2-360 c_{4444}+4320) \chi ^5-20 (45 c_{444}^2+2 
c_{446}^2-360 c_{4444}+4320) \chi ^6
\bigg]\,\log\chi\notag \\
&
+\frac{\pi\chi}{6144}\bigg[
-20 (45 c_{444}^2+2 c_{446}^2-360 c_{4444}+4320)\notag \\
& -10 (45 c_{444}^2+2 c_{446}^2-360 c_{4444}+4320) \chi -36 (21 c_{444}^2-88 c_{4444}+896) 
\chi ^2\notag \\
& +(-909 c_{444}^2+10 c_{446}^2+2952 c_{4444}-26784) \chi ^3-36 
(21 c_{444}^2-88 c_{4444}+896) \chi ^4\notag \\
& -10 (45 c_{444}^2+2 
c_{446}^2-360 c_{4444}+4320) \chi ^5-20 (45 c_{444}^2+2 c_{446}^2-360 
c_{4444}+4320) \chi ^6
\bigg]\,\log(1-\chi)\notag \\
&+\frac{\pi\chi^{2}(1-\chi+\chi^{2})^{2}}{36864(1-\chi)^{6}}\,\bigg[
-120 (45 c_{444}^2+2 c_{446}^2-360 c_{4444}+400)\notag \\
& +360 (45 c_{444}^2+2 
c_{446}^2-360 c_{4444}+400) \chi +(-4851 c_{444}^2-140 
c_{446}^2+44208 c_{4444}-72576) \chi ^2\notag \\
& -2 (8649 c_{444}^2+460 
c_{446}^2-63792 c_{4444}+47424) \chi ^3+(-4851 c_{444}^2-140 
c_{446}^2+44208 c_{4444}-72576) \chi ^4\notag \\
& +360 (45 c_{444}^2+2 
c_{446}^2-360 c_{4444}+400) \chi ^5-120 (45 c_{444}^2+2 c_{446}^2-360 
c_{4444}+400) \chi ^6\bigg].
\end{align}
Again, the logarithmic terms $\sim \log\chi, \log(1-\chi)$ vanish using the explicit form of the couplings. This is due to the 
{\em strange} identities
\be
(c_{444})^{2} = \frac{8}{21}\,(-112+11\,c_{4444}),\qquad
(c_{446})^{2} = \frac{600}{7}\,(-14+\,c_{4444}),
\ee
that can be easily checked using (\ref{2.9}). In conclusion, we find
\begin{align}
& G_{4444}^{\rm AdS}(\chi) = \notag \\ 
& \frac{245\,\pi}
{192\,(n-1)(n-2)(n+3)(n+4)}\,\frac{\,\chi^{2}\,(1-\chi+\chi^{2})^{2}}{(1-\chi)^{6}}\,\bigg[
10 (n-2) (n-1) (n+3) (n+4)\,(1-3\chi-3\chi^{5}+\chi^{6})\notag \\
& 
 +9 (2 n^4+8 n^3-39 n^2-94 n+348) \chi ^2(1+\chi^{2}) +2 (7 n^4+28 n^3+176 n^2+296 
n-2532) \chi ^3
\bigg],
\end{align}
that obeys the correct crossing relations
\be
G_{4444}^{\rm AdS}(\chi) = \tfrac{\chi^{8}}{(1-\chi)^{8}}\,G_{4444}^{\rm AdS}(1-\chi), \qquad 
G_{4444}^{\rm AdS}(\chi) = G_{4444}^{\rm AdS}(\tfrac{\chi}{\chi-1}).
\ee

\subsubsection{Mixed boundary correlator $\llangle \varphi_{3}\varphi_{3}\varphi_{4}\varphi_{4}\rrangle$}

Finally, we can evaluate the mixed 4-point function $\llangle \varphi_{3}\varphi_{3}\varphi_{4}\varphi_{4}\rrangle$
by the same methods.  We do not repeat all the steps to get the final result, since they are completely similar to those
for the diagonal 4-point functions. 
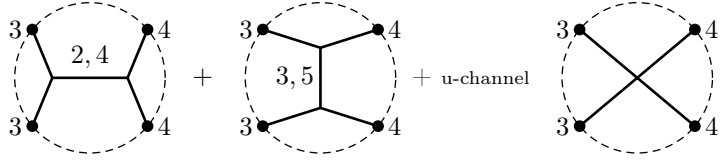
\begin{figure}[ht]
\centering
\begin{tikzpicture}[line width=1 pt, scale=0.5, rotate=0]
\coordinate (A1) at (140:2);
\coordinate (A2) at (-140:2);
\coordinate (A3) at (40:2);
\coordinate (A4) at (-40:2);
\coordinate (M1) at (-1,0);    \coordinate (M2) at (1,0);

\draw[densely dashed,line width=0.5pt] (0,0) circle (2);
\draw (A1)--(M1)--(A2); \draw (A3)--(M2)--(A4); \draw (M1)--(M2);

\draw[fill=black] (A1) circle (0.12); \node[left] at (A1) {$3$};
\draw[fill=black] (A2) circle (0.12); \node[left] at (A2) {$3$};
\draw[fill=black] (A3) circle (0.12); \node[right] at (A3) {$4$};
\draw[fill=black] (A4) circle (0.12); \node[right] at (A4) {$4$};


\node[above] at (0,0) {$2, 4$};
\node at (3,0) {$+$};
\end{tikzpicture}
\begin{tikzpicture}[line width=1 pt, scale=0.5, rotate=0]
\coordinate (A1) at (140:2);
\coordinate (A2) at (-140:2);
\coordinate (A3) at (40:2);
\coordinate (A4) at (-40:2);
\coordinate (M1) at (0,0.8);    \coordinate (M2) at (0,-0.8);

\draw[densely dashed,line width=0.5pt] (0,0) circle (2);
\draw (A1)--(M1)--(A3); \draw (A2)--(M2)--(A4); \draw (M1)--(M2);

\draw[fill=black] (A1) circle (0.12); \node[left] at (A1) {$3$};
\draw[fill=black] (A2) circle (0.12); \node[left] at (A2) {$3$};
\draw[fill=black] (A3) circle (0.12); \node[right] at (A3) {$4$};
\draw[fill=black] (A4) circle (0.12); \node[right] at (A4) {$4$};


\node[left=-0.05cm] at (0,0) {$3, 5$};
\node at (4,0) {+ \scriptsize u-channel};
\end{tikzpicture}
\begin{tikzpicture}[line width=1 pt, scale=0.5, rotate=0]
\coordinate (A1) at (140:2);
\coordinate (A2) at (-140:2);
\coordinate (A3) at (40:2);
\coordinate (A4) at (-40:2);
\coordinate (M1) at (0,0.8);    \coordinate (M2) at (0,-0.8);

\draw[densely dashed,line width=0.5pt] (0,0) circle (2);
\draw (A1)--(0,0)--(A3); \draw (A2)--(0,0)--(A4); 
\draw[fill=black] (A1) circle (0.12); \node[left] at (A1) {$3$};
\draw[fill=black] (A2) circle (0.12); \node[left] at (A2) {$3$};
\draw[fill=black] (A3) circle (0.12); \node[right] at (A3) {$4$};
\draw[fill=black] (A4) circle (0.12); \node[right] at (A4) {$4$};


\end{tikzpicture}
\caption{Tree diagrams contributing the mixed boundary correlator
$\llangle\varphi_{3}\varphi_{3}\varphi_{4}\varphi_{4}\rrangle$.
}\label{fig:3344}
\end{figure}
From the connected diagrams in Fig.~(\ref{fig:3344}) we can write 
\be
\mc C_{3}^{-1}\,\mc C_{4}^{-1}\,
\llangle  \varphi_{3}(t_{1}) \varphi_{3}(t_{2}) \varphi_{4}(t_{3}) \varphi_{4}(t_{4})\rrangle_{\rm conn}= 
\frac{\beta^{2}}{t_{12}^{6}\,t_{34}^{8}}\,G_{3344}^{\rm AdS}(\chi),
\ee
with 
\begin{align}
G_{3344}^{\rm AdS}(\chi) &= 
\frac{105\,\pi}
{128\,(n-1)(n+3)}\,\frac{\chi^{2}}{(1-\chi)^{4}}\,\bigg[
10 (n-1) (n+3) (1-3\chi)+4(n-2)(n+4)\chi^{5}(-8+3\chi)\notag \\
& +(-303+82 n+41 n^{2})\chi^{2}
-8(-57+8n+4n^{2})\,\chi^{3}+(-469+86n+43n^{2})\,\chi^{4}
\bigg],
\end{align}
obeying the single crossing relation
\be
G_{3344}^{\rm AdS}(\chi) = G_{3344}^{\rm AdS}(\tfrac{\chi}{\chi-1}).
\ee


\section{Chiral 4-point functions in $\mc W_{n}$ Virasoro extensions}
\la{sec:w-corr}

We now turn to the CFT side and discuss how to compute the relevant (chiral)  4-point functions 
in the $\mc W_{3}$ and $\mc W_{4}$ extended Virasoro algebras. In particular, we analyze the structure of their large $c$ expansion
and show how to compute the subleading $\mc O(c)$ contribution to the spin 3 and spin 4 correlators
in the generic $\mc W_{n}$ algebra.

\subsection{General structure of 4-point functions in ${\rm 2d}$ CFT}

Let us review some basic facts about the conformal block decomposition of 4-point functions
of primary fields, see for instance \cite{Dolan:2000ut,Dolan:2003hv,Perlmutter:2015iya}.
Let us consider four chiral primaries with 
dimensions $\Delta_{i}$ and define $G(z)$ by 
\be
\la{3.1}
\langle \mc O_{\de_{1}}(\infty)\,\mc O_{\de_{2}}(1)\,\mc O_{\de_{3}}(z)\,\mc O_{\de_{4}}(0)\rangle = 
\lim_{w\to\infty}w^{2\,(\de_{1}+\de_{2})} \langle \mc O_{\de_{1}}(w)\,\mc O_{\de_{2}}(1)\,
\mc O_{\de_{3}}(z)\,\mc O_{\de_{4}}(0)\rangle = \frac{1}{z^{2\,(\de_{3}+\de_{4})}}\,G(z).
\ee
The variable $z$ is again the cross ratio  $\chi=\frac{(z_{1}-z_{2})(z_{3}-z_{4})}{(z_{1}-z_{3})(z_{2}-z_{4})}$ for the 
configuration $z_{i}=(\infty, 1, z, 0)$. As we discussed in the Introduction, 
the reason why we consider chiral primaries is that we want to compare
with the 1d boundary of AdS$_{2}$ that will be parametrized by $z$ restricted to be real $z=\overline z$.

The function $G(z)$ in (\ref{3.1}) may be expanded in the $s$-channel by summing over the 
exchanged primaries $\mc O_{p}$
\be
\la{3.2}
G(z) = \sum_{p}C_{12p}C_{34p}\,\mc F(\bm{\de}, \de_{p}; z),
\ee
where $C_{abc}$ are the three point function coefficients for primaries with 
unit normalization of 2-point function, and an obvious dependence on the central charge is understood.
The Virasoro conformal block $\mc F(\bm{\de}, \de_{p}; z)$ is fully determined by Virasoro symmetry.
It is convenient to present it as a sum over contributions from the level $q$ quasi-primaries appearing in the 
Verma module $M(\Delta_{p})$ built upon $\mc O_{p}$. One has 
\be
\la{3.3}
\mc F(\bm{\de}, \de_{p}; z) = z^{\de_{p}}\,\sum_{q=0}^{\infty}\chi_{q}(\bm{\de}, \de_{p})\,z^{q}\,
_{2}F_{1}(\de_{p}+q+\de_{12}, \de_{p}+q+\de_{34}, 2\,(\de_{p}+q); z),\quad \de_{ij}=\de_{i}-\de_{j},
\ee
where the expansion coefficients $\chi_{q}(\bm{\de}, \de_{p})$
are fully determined by the Virasoro algebra, {\em i.e.} they can be computed by summing over
the explicit quasi-primaries that appear in the Virasoro Verma module built on $\mc O_{-\de_{p}}|0\rangle$. One
important special case is the contribution from the identity $\mc O_{p}=\mathbb{I}$. In this case,
we have $\de_{p}\to 0$ and the constraint $\de_{1}=\de_{2}$ and $\de_{3}=\de_{4}$. Then,
\be
\mc F(\bm{\de},0; z) = \sum_{q=0}^{\infty}\chi_{q}(\bm{\de}, 0)\,z^{q}\,
_{2}F_{1}(q, q, 2q; z),
\ee
with
\begin{align}
\la{3.5}
\chi_{0}(\bm{\de}, 0) &=1, \quad 
\chi_{2}(\bm{\de}, 0)=\frac{2\,\de_{1}\,\de_{3}}{c}, \quad 
\chi_{4}(\bm{\de}, 0)=\frac{10\,(\de_{1}^{2}+\frac{\de_{1}}{5})
\,(\de_{3}^{2}+\frac{\de_{3}}{5})}{c\,(5c+22)},  \notag \\
\chi_{6}(\bm{\de},0) &= \frac{(14\de_{1}^{2}+\de_{1})(14\de_{3}^{2}+\de_{3})}{63\,c\,(70c+29)}\notag \\
&+\frac{4\de_{1}\de_{3}[c(70\de_{1}^{2}+42\de_{1}+8)+29\de_{1}^{2}-57\de_{1}-2]
[c(70\de_{3}^{2}+42\de_{3}+8)+29\de_{3}^{2}-57\de_{3}-2]}
{3\,c\,(2c-1)\,(5c+22)\,(7c+68)\,(70c+29)},
\end{align}
and so forth.
Remarkably, one has the general result
\be
\lim_{c\to\infty} \mc F(\bm{\de}, \de_{p}; z) = z^{\de_{p}}\,_{2}F_{1}(\de_{p}+\de_{12}, \de_{p}
+\de_{34}, 2\de_{p}; z),
\ee
{\em i.e.} the Virasoro block reduces to the global block. This fact will have a role in the following 
discussion of the 4-point function analysis as a boundary 1d CFT, but we shall need some refinement, see 
later (\ref{3.23}).

\subsection{Virasoro extensions and $\mc W_{n}$}

We are interested in CFT with extended Virasoro symmetry associated with 
additional chiral generator $Q_{s}$ with integer spin $s\ge 3$
\cite{Zamolodchikov:1985wn}. This means that we have the singular operator product expansions (OPE)
\be
T(z) T(0) \sim \frac{c}{2\,z^{4}}+\frac{2\,T(0)}{z^{2}}+\frac{T'(0)}{z},\qquad
T(z)Q_{s}(0) \sim \frac{s}{z^{2}}\,Q_{s}(0)+\frac{1}{z}\,Q'_{s}(0),
\ee
where $T$ is the stress-energy tensor, and $c$ the central charge. The conformal Ward identities
read
{\small
\begin{align}
\la{3.8}
& \langle T(z_{1})T(z_{2})\cdots T(z_{N})\,Q_{s_{1}}(w_{1})\cdots Q_{s_{M}}(w_{M})\rangle \notag \\
& = \sum_{i=2}^{N}\frac{c}{2\,(z_{1}-z_{i})^{4}}\,\langle
T(z_{2})\cdots T(z_{i-1})\,T(z_{i+1})\cdots T(z_{N})\,Q_{s_{1}}(w_{1})\cdots Q_{s_{M}}(w_{M})\rangle\notag \\
&+\bigg\{
\sum_{i=2}^{N}\bigg[
\frac{2}{(z_{1}-z_{i})^{2}}+\frac{1}{z_{1}-z_{i}}\frac{\partial}{\partial z_{i}}
\bigg]+\sum_{j=1}^{M}\bigg[
\frac{s_{j}}{(z_{1}-w_{j})^{2}}+\frac{1}{z_{1}-w_{j}}\frac{\partial}{\partial w_{j}}
\bigg]
\bigg\}\,\langle T(z_{2})\cdots T(z_{N})\,Q_{s_{1}}(w_{1})\cdots Q_{s_{M}}(w_{M})\rangle.
\end{align}
}
Without higher spin fields, they imply the well known $T$ correlators
\begin{align}
\la{3.9}
\langle T(z_{1})T(z_{2})\rangle &= \frac{c}{2\,z_{12}^{4}}, \\
\la{3.10}
\langle T(z_{1})T(z_{2})T(z_{3})\rangle &= \sum_{i=2}^{3}\bigg[
\frac{2}{(z_{1}-z_{i})^{2}}+\frac{1}{z_{1}-z_{i}}\frac{\partial}{\partial z_{i}}
\bigg]\,\langle T(z_{2})T(z_{3})\rangle = \frac{c}{z_{12}^{2}\,z_{13}^{2}\,z_{23}^{2}},\notag \\
\langle T(z_{1}) T(z_{2}) T(z_{3}) T(z_{4})\rangle &= \frac{c^{2}}{4}\,\bigg(\frac{1}{z_{12}^{4}\,z_{34}^{4}}
+\frac{1}{z_{13}^{4}\,z_{24}^{4}}+\frac{1}{z_{14}^{4}\,z_{23}^{4}}\bigg)\notag \\
&+c\,\bigg(
\frac{1}{z_{12}^{2}\,z_{23}^{2}\,z_{34}^{2}\,z_{14}^{2}}
+\frac{1}{z_{13}^{2}\,z_{24}^{2}\,z_{14}^{2}\,z_{23}^{2}}
+\frac{1}{z_{12}^{2}\,z_{24}^{2}\,z_{34}^{2}\,z_{13}^{2}}
\bigg).
\end{align}
The simplest case with higher spin fields is $\langle T(z_{1}) Q_{s}(z_{2}) Q_{s}(z_{3})\rangle$.
Assuming the standard (Zamolodchikov) normalization $\langle Q_{s}(z) Q_{s}(0)\rangle = \frac{c}{s}\,\frac{1}{z^{2s}}$,
we obtain, using (\ref{3.8}), the expression, similar to (\ref{3.10}),
\be
\la{3.11}
\langle T(z_{1})Q_{s}(z_{2})Q_{s}(z_{3})\rangle = 
\frac{c}{z_{12}^{2}\,z_{13}^{2}\,z_{23}^{2s-2}}.
\ee
For an extension with spins $s', s'', \dots$, usually denoted by $\mc W(2,s,s'',\dots)$, one can postulate 
a set of OPEs for all the generators (including the stress-tensor). To be consistent, they should be 
equivalent to associativity in correlators, or Jacobi identities. This is by far a non-trivial constraint, for a review of some 
basic constructions see for instance \cite{Bowcock:1991zk,Bouwknegt:1992wg}. Generally speaking, there are 
classes of solutions valid for generic central charge (apart from isolated special singular values) as well
as specific solutions that are valid only at certain central charges. Here we consider the simplest case of the former type, 
{\em i.e.} the so-called quantum $\mc W_{n}\equiv \mc WA_{n-1}$ algebra that is 
the simplest example of a Casimir algebra 
\cite{Fateev:1987vh,Fateev:1987zh,Bais:1987dc,Bais:1987zk} and 
contains higher spin generators $Q_{s}$ with spin $s=3, 4, \dots, n$. As a preliminary step, 
we now discuss in some details the 4-point functions for spin 3 and spin 4 generators
in $\mc W_{3}$ and $\mc W_{4}$. Then, we discuss their large $c$ limit, and its 
generalization to all $\mc W_{n}$.


\subsubsection{4-point functions in the $\mc W_{3}$ algebra}

The simplest Virasoro extension is $\mc W_{3}\equiv\mc W(2,3)$ first discussed in 
\cite{Zamolodchikov:1985wn}. Denoting by $Q_{3}$
the spin-3 primary, we have the fusion data (singular OPE between local conformal families)
\be
\la{3.12}
Q_{3}\,Q_{3} = \frac{c}{3}\, [\mathbb{I}],
\ee
where $[\mathbb{I}]$ is the conformal family of the identity operator. 
Making explicit the descendents, this means the following singular OPE
\be
Q_{3}(z)\, Q_{3}(0) = \frac{c}{3\,z^{6}}+\frac{2T(0)}{z^{4}}+\frac{T'(0)}{z^{3}}
+\frac{1}{z^{2}}\bigg[\tfrac{3}{10}T''(0)+\frac{32}{22+5c}\,\Lambda(0)\bigg]+
\frac{1}{z}\bigg[\tfrac{1}{15}\,T'''(0)+\frac{16}{22+5c}\,\Lambda'(0)\bigg]+\cdots\,,
\ee
where $\Lambda(z)$ is the quasi-primary
composite operator $\Lambda = (TT)-\frac{3}{10}T''$. \footnote{Here and later,
we shall denote by $(\cdots)$ the (conformally) normal ordered composite operators, see for instance
\cite{Bouwknegt:1992wg}. There should not be confusion with ordinary brackets.} Isolating the poles from the OPE
in the 4-point function $\langle Q_{3}Q_{3}Q_{3}Q_{3}\rangle$, we get the exact result
for the $G$-function in (\ref{3.1})
\begin{align}
\la{3.14}
G_{3333}(z) &= \frac{c^{2}}{9}\bigg[1+z^{6}+\frac{z^{6}}{(1-z)^{6}}\bigg]+c\,\bigg[
2 z^4+2 z^3+\frac{9 z^2}{5}+\frac{8 z}{5}-\frac{96}{5 (1-z)}+\frac{99}{5
   (1-z)^2}\notag \\
   & -\frac{10}{(1-z)^3}+\frac{2}{(1-z)^4}+\frac{37}{5}\bigg]
   +\frac{512\,c}{5\,(22+5\,c)}\,\frac{z^{4}}{(1-z)^{2}}.
\end{align}
As a check, one can verify the correct crossing relations
\be
G_{3333}(z) = G_{3333}(\tfrac{z}{z-1}) = \tfrac{z^{6}}{(1-z)^{6}}\,G_{3333}(1-z).
\ee

\subsubsection{4-point functions in the  $\mc W_{4}$ algebra}

The $\mc W_{4}\equiv\mc W(2,3,4)$ algebra fusion rules are discussed, {\em e.g.} in \cite{Kausch:1990bn}. 
Denoting by $Q_{3}$ and $Q_{4}$ the spin-3 and 4 primaries, we have the OPEs
\begin{align}
\la{3.16}
Q_{3}\,Q_{3} &= \frac{c}{3}\,[\mathbb{I}]+\gamma\,[Q_{4}], \notag \\
Q_{3}\,Q_{4} &= \frac{3}{4}\,\gamma\,[Q_{3}],\notag \\
Q_{4}\,Q_{4} &= \frac{c}{4}\,[\mathbb{I}]+\mu\,[Q_{4}]+\lambda\,[\Phi_{6}],
\end{align}
where $\Phi_{6} = (Q_{3}Q_{3})+\cdots$ is the dimension 6 (composite) primary appearing in (\ref{A.4}). Up to automorphisms
changing the sign of $\gamma$, 
the constants in (\ref{3.16}) are
\be
\gamma=\pm\frac{4}{3}\,\sqrt\frac{3\,(7c+114)(c+2)}{(5c+22)(c+7)},\ \ \ \ \
\mu = -12\,\frac{c^{2}+c+218}{(5c+22)(c+7)\,\gamma},\ \ \ \ \
\lambda =\frac{45\,(5c+22)}{2\,(7c+114)(c+2)}.
\ee 
The explicit form of $\Phi_{6}$ is 
\begin{align}
\Phi_{6} &= (Q_{3}Q_{3}) +\frac{(5 c+76) \sqrt{\frac{(c+2) (7 
c+114)}{(c+7) (5 c+22)}}}{9 \sqrt{3} (c+24)} \, Q_{4}''
+\frac{88 \sqrt{\frac{(c+2) (7 c+114)}{(c+7) (5 
c+22)}} }{3 \sqrt{3} (c+24)}\,(T\,Q_{4})\notag\\
& +\frac{1504-2 c (67 c+178)}{(2 c-1) (5 c+22) (7 
c+68)}\, (T''\,T)
 -\frac{c (225 c+1978)+776}{2 (2 c-1) (5
c+22) (7 c+68)}\,(T' \,T')\notag \\
& -\frac{16 (191 
c+22) }{3 (2 c-1) (5 c+22) (7 c+68)}\,(TTT) -\frac{(c-8) [5 c (c+12)+4]}{6 (2 
c-1) (5 c+22) (7 c+68)}\, T^{(4)},
\end{align}
with squared norm
\be
\langle \Phi_{6}(z)\Phi_{6}(0)\rangle = \frac{4 (c-1) c (c+2) (c+13) (3 c+116) (7 c+114)}{27 (c+7) (c+24) (2 \
c-1) (7 c+68) } \ 
\frac{1}{z^{12}}.
\ee
We want to compute the $G$-functions associated with the correlators 
\be
\langle Q_{3}Q_{3}Q_{3}Q_{3}\rangle,\qquad
\langle Q_{3}Q_{3}Q_{4}Q_{4}\rangle,\qquad
\langle Q_{4}Q_{4}Q_{4}Q_{4}\rangle.
\ee
A tedious but straighforward calculation gives \footnote{The package \cite{Thielemans:1991uw} is useful for such
computations.}
\begin{align}
\la{3.21}
G_{3333}(z) &=   \frac{c^{2}}{9}\bigg[1+z^{6}+\frac{z^{6}}{(1-z)^{6}}\bigg]+c\,\bigg[
2z^{4}+2z^{3}+\frac{11}{2}z^{2}+\frac{16}{3}z-\frac{80}{3\,(1-z)}+\frac{65}{3\,(1-z)^{2}}\notag \\
& -\frac{10}{(1-z)^{3}}+\frac{2}{(1-z)^{4}}+13\bigg]
+\frac{100\,c}{3\,(c+7)}\,\frac{z^{4}}{(1-z)^{2}},\notag \\
G_{3344}(z) &=   \frac{c^{2}}{12}+c\,\bigg[
\frac{7z^{4}}{5}+\frac{28z^{3}}{15}+2z^{2}+2z+\frac{7}{5(1-z)^{4}}-\frac{112}{15(1-z)^{3}}
+\frac{16}{(1-z)^{2}}-\frac{86}{5(1-z)}+\frac{109}{15}
\bigg]\notag \\
&+\frac{c}{15(7+c)(22+5c)}\,\frac{z^{4}}{(1-z)^{4}}\,[2\,(2844-5688 z+3092 z^{2}-248 z^{3}+93 z^{4})
\notag \\
&+c\,(2484-4968 z+4412 z^{2}-1928 z^{3}+723 z^{4})],\notag \\
G_{4444}(z) &=  \frac{c^{2}}{16}\bigg[1+z^{8}+\frac{z^{8}}{(1-z)^{8}}\bigg]+c\,\bigg[
2z^{6}+2z^{5}+\frac{279z^{4}}{140}+\frac{139z^{3}}{70}+\frac{55z^{2}}{28}+\frac{27z}{14}\notag \\
&+\frac{2}{(1-z)^{6}}-\frac{14}{(1-z)^{5}}+\frac{5879}{140(1-z)^{4}}-\frac{4897}{70(1-z)^{3}}
+\frac{9783}{140(1-z)^{2}}-\frac{585}{14(1-z)}+\frac{831}{70}\bigg]\notag \\
&+
\frac{9\,c\,(9264108+3031912 c+503031 c^{2}+16301 c^{3})}{140(2+c)(7+c)(22+5c)(114+7c)}
\frac{z^{4}(1-z+z^{2})^{2}}{(1-z)^{4}}.
\end{align}
These obey the exact crossing relations
\begin{align}
G_{3333}(z) &= G_{3333}(\tfrac{z}{z-1}) = \tfrac{z^{6}}{(1-z)^{6}}\,G_{3333}(1-z), &  
G_{4444}(z) = G_{4444}(\tfrac{z}{z-1}) = \tfrac{z^{8}}{(1-z)^{8}}\,G_{4444}(1-z),\notag \\ 
G_{3344}(z) &= G_{3344}(\tfrac{z}{z-1}).
\end{align}

\subsection{Large $c$ analysis}

The $G$-functions in (\ref{3.14})  and (\ref{3.21}) may be expanded at large central charge
\be
\la{3.23}
G(z) = c^{2}\,G_{0}(z)+c\,G_{1}(z) + \mc O(c^{0}). 
\ee
The $\mc O(c^{2})$ contributions are obvious, they come from {\em disconnected} contributions where two pairs of fields
fuse into the identity. The next-to-leading $\mc O(c)$ terms, {\em i.e.} $G_{1}(z)$,
display a certain regularity and structural similarity. Finally, the NNLO contributions $\mc O(c^{0})$
appear to be more involved, but for our present purposes we are interested in the  leading and next-to-leading terms. 

\medskip
\noindent
It is useful to analyze the small $z$ expansion of the $G$ functions in terms of conformal blocks
at large $c$. The wealth of data we want to reproduce is summarized by the following expansions
-- we add an algebra suffix for better clarity -- 
\begin{align}
\la{3.24}
G_{3333}^{\mc W_{3}}(z) &= c^2\, (\tfrac{1}{9}+\tfrac{2}{9}z^{6}+\tfrac{2}{3}z^{7}
+\tfrac{7}{3}z^{8}+\cdots)+c\, (2 z^2+2 z^3+\tfrac{9}{5}z^4+\tfrac{8}{5} z^5
+\tfrac{37}{5} z^6+\tfrac{96}{5} z^6+39 z^8+\cdots)+\mc O(c^{0}), \notag \\
G_{3333}^{\mc W_{4}}(z) &= c^2 \, (\tfrac{1}{9}+\tfrac{2}{9}z^{6}+\tfrac{2}{3}z^{7}
+\tfrac{7}{3}z^{8}+\cdots)+c\, (2 z^2+2 z^3+\tfrac{11}{3}z^4+\tfrac{16}{3} z^5+13 z^6
+\tfrac{80}{3}z^{7}+\tfrac{145}{3}z^{8}+\cdots)+\mc O(c^{0}), \notag \\
G_{3344}^{\mc W_{4}}(z) &= \tfrac{c^2}{12}+c (2 z^2+2 z^3+\tfrac{6}{5} z^4+\tfrac{2}{5} z^5
+\tfrac{10}{3} z^6+10 z^7+\tfrac{109}{5} z^8+\cdots)+\mc O(c^{0}), \notag \\
G_{4444}^{\mc W_{4}}(z) &= c^2\, (\tfrac{1}{16}+\tfrac{1}{8}z^{8}+\cdots)+c\, 
(2 z^2+2 z^3+\tfrac{279}{140}z^{4}+\tfrac{139}{70}z^5+\tfrac{55}{28}z^6+\tfrac{27}{14} z^7
+\tfrac{831}{70}z^8+\cdots)+\mc O(c^{0}).
\end{align}
In order to match (\ref{3.24}) and  the general representation (\ref{3.2}), we need 
the large $c$ expansion of Virasoro blocks. Let us  focus on the case
\be
\de_{1}=\de_{2}=\de,\qquad \de_{3}=\de_{4}=\de'.
\ee
A systematic expansion of the Virasoro conformal block at large $c$ and fixed dimensions $\bm{\de}$, $\de_{p}$
has been computed in \cite{Fitzpatrick:2016mtp,Hikida:2017ehf,Hikida:2018dxe}. The conformal block can be written 
\begin{align}
\la{3.26}
\mc F(\de, \de', \de_{p}; z) &= z^{\de_{p}}\,\bigg[\, F_{0}(\de_{p})
+\frac{1}{c}\,F_{1}(\de, \de' ; \de_{p})
+\cdots\bigg],\notag \\
F_{0}(\de_{p}) &= {_{2}}F_{1}(\de_{p}, \de_{p}, 2\,\de_{p}; z), \notag \\
F_{1}(\de, \de'; \de_{p}) &= 12\,\bigg[\mathsf{f}_{a}(\de_{p})\,\de\,\de'
+\mathsf{f}_{b}(\de_{p})\,(\de+\de')+\mathsf{f}_{c}(\de_{p})
\bigg],
\end{align}
where the explicit functions $\mathsf{f}_{a}(\de_{p})$ may be found in convenient form in \cite{Bombini:2018jrg}.
In particular, for the vacuum block $\de_{p}=0$ one has simply
\be
\la{3.27}
F_{1}(\de, \de'; 0) = -12\,\de\,\de'\,\bigg[2+\frac{2-z}{z}\log(1-z)\bigg]. 
\ee
%
%
%
Let us now discuss the various cases in (\ref{3.24}) from this perspective and by means of these tools.

\subsubsection{The $\mc W_{3}$ case}

We begin with a finite $c$ analysis.  The simple fusion algebra (\ref{3.12}) implies that $G^{\mc W_{3}}_{3333}(z)$ 
starts with the vacuum 
block, {\em cf.} (\ref{3.2}), and continues with other primary contributions that  belong to the 
regular part of the OPE. So we expect 
\be
\la{3.28}
G^{\mc W3}_{3333}(z) = \frac{c^{2}}{9}\,\mc F(\{3,3,3,3\}, 0; z)+\text{other primary contributions}.
\ee
The first primary is $\Phi_{6}= (Q_{3}Q_{3})+\cdots$ and  has dimension 6, {\em cf.} (\ref{A.3}). 
The explicit form of this primary, normalized in order to have unit 2-point 
function, is 
\begin{align}
\Phi_{6} &= \frac{3}{2} \sqrt{\frac{(2 c-1) (5 c+22) (7 c+68)}{c (c+2) (c+23) (5 
c-4) (7 c+114)}}\,\bigg[ 
(Q_{3}Q_{3})
+\frac{1504-2 c (67 c+178)}{(2 c-1) (5 c+22) (7 c+68)}\, (T''\,T)\notag \\
& -\frac{c (225 c+1978)+776 }{2 (2 c-1) (5 c+22) (7 c+68)}\,(T'\,T')
 -\frac{16 (191 c+22) }{3 (2 c-1) (5 c+22) (7 c+68)}\,(T,(T,T))\notag \\
& -\frac{(c-8) [5 c (c+12)+4]}{6 (2 c-1) (5 c+22) (7 c+68)}\,T^{(4)}
\bigg].
\end{align}
This is fully consistent with (\ref{3.2}). Indeed, from (\ref{3.14}) we can write 
\be
G^{\mc W_{3}}_{3333}(z) = \frac{c^{2}}{9}\bigg[\mathsf{F}_{0}(z)+\frac{18}{c}\,\mathsf{F}_{2}(z)
+\frac{4608}{5\,c\,(22+5\,c)}\,\mathsf{F}_{4}(z)+\frac{9710+2189\,c+70\,c^{2}}{7\,c\,(22+5\,c)}\,
\mathsf{F}_{6}(z)+\mc O(z^{8})\bigg],
\ee
where $\mathsf{F}_{q}(z)=z^{q}\,_{2}F_{1}(q,q,2q; z)$. Comparing 
the coefficients of the hypergeometric functions  with (\ref{3.5}) at $\de_{1}=\de_{3}=3$
we see that we can continue (\ref{3.28}) as 
\be
\la{3.31}
G^{\mc W_{3}}_{3333}(z) = \frac{c^{2}}{9}\,\mc F(\{3,3,3,3\}, 0; z)+
\frac{4 \,c\,(c+2) (c+23) (5 c-4) (7 c+114)}{9(2 c-1) (5 c+22) (7 c+68)}\,\mc F(\{3,3,3,3\}, 6; z)+\mc O(z^{8}).
\ee
The coefficients are in agreement with (\ref{3.2}) taking into account that 
$\langle Q_{3}(z_{1})Q_{3}(z_{2})\rangle = \frac{c^{2}}{9\,z_{12}^{6}}$, and that the regular part
of the OPE $Q_{3}(z)Q_{3}(0)$ starts with $(Q_{3}Q_{3})+\cdots$ .

\medskip
\noindent
The large $c$ limit of  (\ref{3.31}) may be computed by expanding both the
coefficients and the conformal blocks. This gives
\be
\la{3.32}
G^{\mc W_{3}}_{3333}(z) = \frac{c^{2}}{9}\,\bigg[1+\frac{1}{c}\,F_{1}(3,3; 0)+\cdots\bigg]
+\bigg[\frac{2c^{2}}{9}+\frac{209\,c}{35}+\cdots\bigg]\,z^{6}\,
\bigg[F_{0}(6)+\frac{1}{c}\,F_{1}(3,3,6)+\cdots\bigg]+\mc O(z^{8}).
\ee
Using the explicit expressions 
\begin{align}
\la{3.33}
F_{1}&(3,3; 0) = 108 \left[-\frac{(2-z) \log (1-z)}{z}-2\right] = 
18 z^2+18 z^3+\frac{81 z^4}{5}+\frac{72 z^5}{5}+\frac{90 
z^6}{7}+\frac{81 z^7}{7}+\frac{21 z^8}{2}+\cdots, \notag \\
F_{1}&(3,3; 6) = \frac{997920}{z^{11}} (z-2) (z^4-28 z^3+154 z^2-252 z+126) 
\text{Li}_2(z)\notag \\
&-\frac{36}{5 z^{11}} (z-2) (113207 z^4-2634800 z^3+13715240 
z^2-22160880 z+11080440) \log (1-z)\notag \\
& +\frac{99792}{z^{12}}
(8 z^6-306 z^5+2835 z^4-10640 z^3+18900 z^2-15876 z+5082) \log 
^2(1-z)\notag \\
& -\frac{12}{z^{10}} (157999 
z^4-2511894 z^3+10520958 z^2-16018128 z+8009064) \notag \\
&= \frac{5832 z^2}{169}+\frac{23328 z^3}{169}+\frac{556767 
z^4}{1690}+\frac{206847 z^5}{338}+\frac{955529751 \
z^6}{976820}+\frac{345575934 z^7}{244205}+\cdots\ ,
\end{align}
one indeed checks that (\ref{3.32}) reads
\be
G^{\mc W_{3}}_{3333}(z) = c^2 [\tfrac{1}{9}+\tfrac{2}{9} z^6+\tfrac{2}{3} z^7+\mc O(z^8)]
+c \,[2 z^2+2 
z^3+\tfrac{9}{5} z^4+\tfrac{8}{5} z^5+\tfrac{37}{5} z^6+\tfrac{96}{5} z^7+\mc O(z^8)]+\mc O(c^{0}),
\ee
in agreement with (\ref{3.24}). The  $\mc O(c)$ contribution, has a very non-trivial 
origin. They depend on the $\mc O(c)$ term of the complicated coefficient in (\ref{3.31}). Besides, looking at the
$\mc O(z^{8})$ term in $G(z)$ and in the representation limited to the two terms in (\ref{3.31}) one sees that there are 
contributions associated with the dimension 8 primary built with $Q_{3}$ and $T$.

\paragraph{Exploiting the analytic structure} At this point, let us make a simple but useful remark. At order 
$\mc O(z^{5})$, the full contribution to $G^{\mc W_{3}}_{3333, 1}(z)$
comes from, {\em cf.} (\ref{3.32}), 
\be
\la{3.35}
G^{\mc W_{3}}_{3333, 1}(z) = \frac{1}{9}F_{1}(3,3; 0) +\mc O(z^{6}) = 
2z^{2}+2z^{3}+\tfrac{9}{5}z^{4}+\tfrac{8}{5}z^{5}+\mc O(z^{6}).
\ee
On the other hand, from crossing symmetry and the meromorphic structure of correlators, we can 
write 
\be
\la{3.36}
G^{\mc W_{3}}_{3333, 1}(z) = P^{\mc W_{3}}(z)+P^{\mc W_{3}}(\tfrac{z}{z-1}),
\qquad P^{\mc W_{3}}(z) = 2\,z^{4}+2\,z^{3}+k_{1}\,z^{2}+k_{2}\,z.
\ee
The second crossing condition $G^{\mc W_{3}}_{3333, 1}(z)=(\tfrac{z}{1-z})^{6}\,
G^{\mc W_{3}}_{3333, 1}(1-z)$ determines $k_{2} = 2\,(k_{1}-1)$. Thus, $P(z)$ has only one free parameter. This 
appears in the small $z$ expansion of (\ref{3.36})
\be
G^{\mc W_{3}}_{3333, 1}(z) = 2\,z^{2}+2\,z^{3}+k_{1}\,z^{4}+2\,(k_{1}-1)\,z^{5}+\mc O(z^{6}).
\ee
Comparing with (\ref{3.35}) we fix $k_{1}=\frac{9}{5}$ and $P^{\mc W_{3}}(z)$ is determined.

\medskip
\noindent
In summary, it has been possible to compute $G^{\mc W_{3}}_{3333, 1}(z)$ by just using the expression for 
$F_{1}(3,3,0)$ and some analytical constraint from the meromorphic structure of the correlators. The representation
(\ref{3.36}) fully captures the exact $\mc O(c)$ term in (\ref{3.21}).

\subsubsection{The $\mc W_{4}$ case}

Let us begin with $G^{\mc W_{4}}_{3333,1}$. Now, the starting point (\ref{3.35}) is not enough because 
the fusion (\ref{3.16}) implies a contribution to (\ref{3.2}) from $Q_{3}\times Q_{3}\to Q_{4}$ at order $\mc O(z^{4})$.
However, this is governed by $\gamma^{2}= \frac{112}{15}+ \mc O(c^{-1})$ at large $c$. This means
\be
\la{3.38}
G^{\mc W_{4}}_{3333, 1}(z) = \tfrac{1}{9}F_{1}(3,3; 0) +\tfrac{1}{4}\,\tfrac{112}{15}F_{0}(4)+\mc O(z^{6}) = 
2z^{2}+2z^{3}+\tfrac{11}{3}z^{4}+\tfrac{16}{3}z^{5}+\mc O(z^{6}).
\ee
Matching this to the representation (\ref{3.36}) and imposing crossing under $z\to 1-z$
is enough to completely determine 
\be
\la{3.39}
G^{\mc W_{4}}_{3333, 1}(z) = P^{\mc W_{4}}_{3333}(z)
+P^{\mc W_{4}}_{3333}(\tfrac{z}{z-1}),\qquad P^{\mc W_{4}}_{3333}(z) = 2\,z^{4}+2\,z^{3}+\tfrac{11}{3}\,z^{2}+
\tfrac{16}{3}\,z.
\ee
The same strategy may be applied to $\langle Q_{3}Q_{3}Q_{4}Q_{4}\rangle$. 
The identity exchange will require the correction
\be
F_{1}(3,4,0) = \tfrac{4}{3}F_{1}(3,4,0).
\ee
Besides, the s-channel exchange of $Q_{4}$
gets a contribution proportional to 
\be
\gamma\mu = -\frac{12}{5}+\mc O(c^{-1}).
\ee
Thus we predict (the factor $\frac{1}{s}=\frac{1}{4}$ is due to the normalization of the 2-point function of 
$Q_{4}$ )
\be
\la{3.42}
G^{\mc W_{4}}_{3344, 1}(z) = \tfrac{1}{12}F_{1}(3,4; 0) -\tfrac{1}{4}\,\tfrac{12}{5}F_{0}(4)+\mc O(z^{5}) = 
2z^{2}+2z^{3}+\tfrac{6}{5}z^{4}+\tfrac{2}{5}z^{5}+\mc O(z^{6}).
\ee
This is not enough to determine the  manifestly crossing invariant polynomial representation because we have 
less symmetry than in the previous case of 4-point functions with equal $\de$'s. However, the fusion $Q_{4}\times Q_{4}\to \Phi_{6}$
has subleading coefficient $\lambda = \frac{225}{14c}+\cdots$ . Taking into account the normalization from the
three point function $\langle Q_{3}Q_{3}(Q_{3}Q_{3})\rangle$, this gives the improved version of (\ref{3.42})
\begin{align}
G^{\mc W_{4}}_{3344, 1}(z) &= \tfrac{1}{12}F_{1}(3,4; 0) -\tfrac{1}{4}\,\tfrac{12}{5}F_{0}(4)
+2\times (\tfrac{1}{3})^{2}\,\tfrac{225}{14}\,F_{0}(6) 
+\mc O(z^{8}) \notag \\
& = 2z^{2}+2z^{3}+\tfrac{6}{5}z^{4}+\tfrac{2}{5}z^{5}+\tfrac{10}{3}z^{6}+10z^{7}+\mc O(z^{8}),
\end{align}
and this is enough to obtain the representation
\be
\la{3.44}
G^{\mc W_{4}}_{3344, 1}(z) = 
P^{\mc W_{4}}_{3344}(z)+P^{\mc W_{4}}_{3344}(\tfrac{z}{z-1}),\qquad 
P^{\mc W_{4}}_{3344}(z) = 2\,z+2\,z^{2}+
\tfrac{28}{15}\,z^{3}+
\tfrac{7}{5}\,z^{4}.
\ee
Alternatively, one can combine the s-channel expansion (\ref{3.42}), with the t-channel expansion $3\times 4\to 3 + \cdots$.
This gives \footnote{In (\ref{3.45}), the factor $\frac{3}{4}\gamma$ is the $Q_{3}Q_{4}\to Q_{3}$ fusion coefficient, see 
(\ref{3.16}). The factor $\frac{1}{3}$ is the inverse spin of the exchanged field, again due to normalization of the two 
point functions. Finally, the $\pm 1$ shift in the $_{2}F_{1}$ arguments are the conformal dimension difference, see (\ref{3.3}).}
\be
\la{3.45}
z^{3}\,G^{\mc W_{4}}_{3344, 1}(z^{-1}) = \left(\tfrac{3}{4}\right)^{2}\,
\tfrac{112}{5}\,\tfrac{1}{3}\,z^{3}\,_{2}F_{1}(3+1,3-1,6,z)+\mc O(z^{5})
= \tfrac{7}{5}\,z^{3}+\tfrac{28}{15}\,z^{4}+\mc O(z^{5}),
\ee
and the combination of (\ref{3.42}) and (\ref{3.45}) fully determine the polynomial $P(z)$ and agrees with (\ref{3.44}).

Finally, the $\langle Q_{4}Q_{4}Q_{4}Q_{4}\rangle$ 4-point function is rather simple.  Its polynomial
representation obeying all crossing constraints is 
\begin{align}
G^{\mc W_{4}}_{4444, 1}(z) &= P^{\mc W_{4}}_{4444}(z)+P^{\mc W_{4}}_{4444}(\tfrac{z}{z-1}), \notag \\
P^{\mc W_{4}}_{4444}(z) &= 2\,z^{6}+2\,z^{5}+ k\,z^{4}+2\,(k-1)\,z^{3}+(5\,k-8)\,z^{2}+2\,(5k-9)\,z.
\end{align}
Thus, we just need to determine $k$ that appears already at order $\mc O(z^{4})$
\be
G^{\mc W_{4}}_{4444, 1}(z)  = 2z^{2}+2z^{3}+k\,z^{4}+2(k-1)\,z^{5}+(5k-8)\,z^{6}+\mc O(z^{7}).
\ee
On the other hand, using $\mu^{2} = \frac{27}{35}+\mc O(c^{-1})$, we can certainly write
\begin{align}
\la{3.48}
G^{\mc W_{4}}_{4444, 1}(z) &= \tfrac{1}{16}F_{1}(4,4; 0) +\tfrac{1}{4}\,\tfrac{27}{35}F_{0}(4)+
\mc O(z^{5}) = 2z^{2}+2z^{3}+\tfrac{279}{140}z^{4}+\cdots.
\end{align}
This fixes $k=\frac{279}{140}$ and determines
\be
P^{\mc W_{4}}_{4444}(z) = 2 z^6+2 z^5+\tfrac{279}{140} z^4+\tfrac{139}{70}z^{3}+\tfrac{55}{28} z^2+\tfrac{27}{14} z.
\ee

\subsection{Computing the 4-point functions in $\mc W_{n}$}

We have analyzed the $\mc W_{3}$ and $\mc W_{4}$ cases to understand
what is the origin of the $c\to \infty$ subleading contribution   to the 4-point functions of spin 3 and 4 generators. 
This is important to generalize the derivation to $\mc W_{n}$. 
We have shown that the diagonal 4-point functions $\langle Q_{3}Q_{3}Q_{3}Q_{3}\rangle$ and 
$\langle Q_{4}Q_{4}Q_{4}Q_{4}\rangle$
may be computed at order $\mc O(c)$
in terms of the large $c$ expansion of the couplings $\gamma, \mu$
in the $Q_{3}Q_{3}$ and $Q_{4}Q_{4}$ OPEs. Other primaries may be present in the 
$Q_{4}Q_{4}$ OPE, but they do not enter our method of calculation.
Instead, in the mixed 4-point function $\langle Q_{3}Q_{3}Q_{4}Q_{4}\rangle$ we needed more information, and in particular 
the primary structure at dimension 6, including the coupling $\lambda$. Nevertheless we have seen that 
by combining the conformal block expansions in the $s$- and $t$-channels,  these problems
can be overcome.

\medskip
\noindent
The above considerations are enough to compute the 4-point functions of spin 3 and 4 in the 
extended $\mc W_{n}$ algebra. To this aim, we 
just require the $n$-dependent values of the couplings $\gamma\to \gamma_{n}$ and $\mu\to \mu_{n}$. 
These have been computed 
in \cite{Hornfeck:1992he} (see also \cite{Hornfeck:1993kp,Blumenhagen:1994wg}) 
based on the free field representation derived in \cite{Fateev:1987zh}.
In our notation, we have the following couplings in $\mc W_{n}$  \footnote{
In a more modern perspective, the couplings in (\ref{3.50}) should be thought as a special limit
of the structure constants of the quantum algebra $\mc W_{\infty}[\nu]$ when $\nu=n$, see also 
\cite{Linshaw:2017tvv}. 
They 
are known to obey a remarkable triality  symmetry with respect to the $\nu$ parameter \cite{Gaberdiel:2012ku}.
}
\begin{align}
\la{3.50}
(\gamma_{n})^{2} &= 64\,\frac{n-3}{n-2}\,\frac{c+2}{5c+22}\,
\frac{c\,(n+3)+2\,(4n+3)\,(n-1)}{c\,(n+2)+(3n+2)\,(n-1)},\notag \\
\mu_{n}\,\gamma_{n} &= \frac{48}{n-2}\,
\frac{c^{2}(n^{2}-19)+3c(6n^{3}-25n^{2}+15)+2(n-1)(6n^{2}-41n-41)}
{(5c+22)[c(n+2)+(3n+2)(n-1)]}.
\end{align}
In particular, expanding at large central charge, 
\be
(\gamma_{n})^{2} = \frac{64}{5}\,\frac{n^{2}-9}{n^{2}-4}+
\mc O(c^{-1}),\qquad
(\mu_{n})^{2} = \frac{36}{5}\,\frac{(n^{2}-19)^{2}}{(n^{2}-4)(n^{2}-9)}+\mc O(c^{-1}).
\ee
The same calculation we did in the $\mc W_{4}$ case, {\em cf.} (\ref{3.38}) and (\ref{3.48}),
gives now the general $\langle Q_{3}Q_{3}Q_{3}Q_{3}\rangle$ 4-point function (at order $\mc O(c)$) in terms of the 
polynomial
\begin{align}
\la{3.52}
& P^{\mc W_{n}}_{3333}(z) = 2z^{4}+2z^{3}+
\frac{5 n^2-36}{(n-2) (n+2)} z^2+\frac{8 (n^2-8)}{(n-2) (n+2)}z.
\end{align}
It reads 
\begin{align}
\la{3.53}
 G_{3333,1}^{\mc W_{n}}(z) &= P^{\mc W_{n}}_{3333}(z)+P^{\mc W_{n}}_{3333}(\tfrac{z}{z-1}) = 
 \frac{1}{n^{2}-4}\frac{z^{2}}{(1-z)^{4}}\,\bigg[
2\,(n^{2}-4)\,(1-3z-3z^{5}+z^{6})\notag \\
& +(9n^{2}-52)\,z^{2}(1+z^{2})-8\,(n^{2}-8)\,z^{3}\bigg].
\end{align}
The mixed 4-point function $\langle Q_{3}Q_{3}Q_{4}Q_{4}\rangle$
is determined by the polynomial 
\be
P^{\mc W_{n}}_{\rm mix}(z) = 
\frac{12}{5}\,\frac{n^{2}-9}{n^{2}-4} z^{4}
+\frac{16}{5}\,\frac{n^{2}-9}{n^{2}-4} z^{3}
+\frac{7n^{2}-88}{n^{2}-4} z^{2}+
12\,\frac{n^{2}-14}{n^{2}-4} z,
\ee
and reads
\begin{align}
\la{3.55}
& G_{3344,1}^{\mc W_{n}}(z) = P^{\mc W_{n}}_{\rm mix}(z)+P^{\mc W_{n}}_{\rm mix}(\tfrac{z}{z-1})  = 
 \frac{1}{5\,(n^{2}-4)}\frac{z^{2}}{(1-z)^{4}}\, \\
&\qquad \bigg[10\,(n^{2}-4)(1-3z) +(41n^{2}-344)z^{2}-8(4n^{2}-61)z^{3}
+(43n^{2}-512)z^{4}+4(n^{2}-9)z^{5}(-8+3z)
\bigg].\notag 
\end{align}
The  general $\langle Q_{4}Q_{4}Q_{4}Q_{4}\rangle$ 4-point function turns out to be expressed in terms
of the polynomial
\begin{align}
P^{\mc W_{n}}_{4444}(z) &=  
2 z^6
+2 z^5
+\frac{9 (2 n^4-51 n^2+397)}{5 (n^2-9) (n^2-4)}  z^4
+\frac{2 (13 n^4-394 n^2+3393) }{5 (n^2-9) (n^2-4)} z^{3}\notag \\
& \ \ \ 
+\frac{5 (2 n^4-71 n^2+657) }{(n^2-9) (n^2-4)} z^{2}
+\frac{18 (n^2-19)^2 }{(n^2-9) (n^2-4)} z,
\end{align}
and reads \footnote{
We remark that the final expressions (\ref{3.53}, \ref{3.55}, \ref{3.57}) have a finite limit for $n\to \infty$.}
\begin{align}
\la{3.57}
 G_{4444,1}^{\mc W_{n}}(z) &= P^{\mc W_{n}}_{4444}(z)+P^{\mc W_{n}}_{4444}(\tfrac{z}{z-1})  = 
 \frac{1}{5\,(n^{2}-4)(n^{2}-9)}\frac{z^{2}(1-z+z^{2})^{2}}{(1-z)^{6}}\,\notag \\
&\bigg[10\,(n^{2}-4)(n^{2}-9)(1-3z-3z^{5}+z^{6}) +9\,(397-51n^{2}+2n^{4})\,z^{2}\,(1+z^{2})
\notag \\
& +2\,(-2673+134n^{2}+7n^{4})\,z^{3}\bigg].
\end{align}

\section{Matching the two sides of the correspondence}
\la{sec:final}

We now have all the ingredients needed to compare the small $\beta$ limit on AdS with the 
large $c$ limit on the CFT side. 

\subsection{Field/generators normalization}
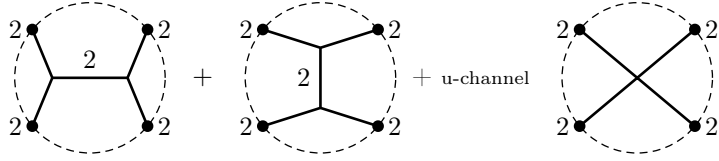
\begin{figure}[ht]
\centering
\begin{tikzpicture}[line width=1 pt, scale=0.5, rotate=0]
\coordinate (A1) at (140:2);
\coordinate (A2) at (-140:2);
\coordinate (A3) at (40:2);
\coordinate (A4) at (-40:2);
\coordinate (M1) at (-1,0);    \coordinate (M2) at (1,0);

\draw[densely dashed,line width=0.5pt] (0,0) circle (2);
\draw (A1)--(M1)--(A2); \draw (A3)--(M2)--(A4); \draw (M1)--(M2);

\draw[fill=black] (A1) circle (0.12); \node[left] at (A1) {$2$};
\draw[fill=black] (A2) circle (0.12); \node[left] at (A2) {$2$};
\draw[fill=black] (A3) circle (0.12); \node[right] at (A3) {$2$};
\draw[fill=black] (A4) circle (0.12); \node[right] at (A4) {$2$};


\node[above] at (0,0) {$2$};
\node at (3,0) {$+$};
\end{tikzpicture}
\begin{tikzpicture}[line width=1 pt, scale=0.5, rotate=0]
\coordinate (A1) at (140:2);
\coordinate (A2) at (-140:2);
\coordinate (A3) at (40:2);
\coordinate (A4) at (-40:2);
\coordinate (M1) at (0,0.8);    \coordinate (M2) at (0,-0.8);

\draw[densely dashed,line width=0.5pt] (0,0) circle (2);
\draw (A1)--(M1)--(A3); \draw (A2)--(M2)--(A4); \draw (M1)--(M2);

\draw[fill=black] (A1) circle (0.12); \node[left] at (A1) {$2$};
\draw[fill=black] (A2) circle (0.12); \node[left] at (A2) {$2$};
\draw[fill=black] (A3) circle (0.12); \node[right] at (A3) {$2$};
\draw[fill=black] (A4) circle (0.12); \node[right] at (A4) {$2$};


\node[left] at (0,0) {$2$};
\node at (4,0) {+ \scriptsize u-channel};
\end{tikzpicture}
\begin{tikzpicture}[line width=1 pt, scale=0.5, rotate=0]
\coordinate (A1) at (140:2);
\coordinate (A2) at (-140:2);
\coordinate (A3) at (40:2);
\coordinate (A4) at (-40:2);
\coordinate (M1) at (0,0.8);    \coordinate (M2) at (0,-0.8);

\draw[densely dashed,line width=0.5pt] (0,0) circle (2);
\draw (A1)--(0,0)--(A3); \draw (A2)--(0,0)--(A4); 
\draw[fill=black] (A1) circle (0.12); \node[left] at (A1) {$2$};
\draw[fill=black] (A2) circle (0.12); \node[left] at (A2) {$2$};
\draw[fill=black] (A3) circle (0.12); \node[right] at (A3) {$2$};
\draw[fill=black] (A4) circle (0.12); \node[right] at (A4) {$2$};


\end{tikzpicture}
\caption{Tree diagrams contributing $\llangle\varphi_{2}\varphi_{2}\varphi_{2}\varphi_{2}\rrangle$.
The external points and internal exchange are labeled by their $\de$.
}\label{fig:A1}
\end{figure}

The matching is based on the  correspondence
\be
\varphi_{s} \to  \kappa_{s}\,Q_{s}.
\ee
The constant $\kappa_{2}$ is somewhat special since $Q_{2}\equiv T$, the stress-energy tensor. We can fix
$\kappa_{2}$ as in \cite{Ouyang:2019xdd} by considering the Liouville $A_{1}$ Toda theory.
The Lagrangian for the $\de=2$ field $\varphi_{2}$ is  
\be
\mc L= \frac{1}{2}\partial^{\mu}\varphi_{2}\partial_{\mu}\varphi_{2}+ \varphi_{2}^2+\frac{2}{3}\,\beta\,
\varphi_{2}^3+\frac{1}{3} \beta ^2 \varphi_{2}^4+\cdots\, .
\ee
At leading order, the constant $\kappa_{2}$ is fixed by looking at the two point function. On AdS, we have -- for our
normalization of the bulk-to-boundary propagator -- 
\be
\llangle \varphi_{2}(t)\varphi_{2}(0)\rrangle = \frac{1+\mc O(\beta^{2})}{t^{4}},
\ee 
and also
\be
\llangle \varphi_{2}(t)\varphi_{2}(0)\rrangle = \kappa_{2}^{2}\,\langle T(t) T(0)\rangle = \frac{\kappa_{2}^{2}\,c}{2}
\,\frac{1}{t^{4}}.
\ee 
Hence (see later for the sign),
\be
\la{4.5}
\kappa_{2} = -\sqrt\frac{2}{c}\,[1+\mc O(\beta^{2})].
\ee
To find a relation connecting $\beta$ with $c$, we need connected diagrams. 
The associated (Witten) Feynman rules are (minus is from $e^{-S}$)
\begin{center}
\begin{tikzpicture}[line width=1 pt, scale=0.6]
\draw (0:1)--(0,0);
\draw (120:1)--(0,0);
\draw (-120:1)--(0,0);
\node[right] at (0:1) {$= -4\,\beta,$};
\end{tikzpicture}
\hskip 2cm
\begin{tikzpicture}[line width=1 pt, scale=0.6]
\draw (45:1)--(0,0);
\draw (135:1)--(0,0);
\draw (225:1)--(0,0);
\draw (315:1)--(0,0);
\node[right] at (0:1) {$= -8\,\beta^{2}.$};
\end{tikzpicture}
\end{center}
Using (\ref{2.13}) we can compute the 3-point function 
\be
\mc C_{2}^{-3/2}\,\llangle \varphi_{2}(t_{1})\varphi_{2}(t_{2})\varphi_{2}(t_{3})\rrangle = (-4\beta)\times \frac{3\,\pi}{8}\,
\frac{1}{t_{12}^{2}\,t_{13}^{2}\,t_{23}^{3}}\,[1+\mc O(\beta^{2})].
\ee
From (\ref{3.10}) we have \footnote{This is standard $6/\beta^{2}$ times $2\pi$ from the 
missing $1/(2\pi)$ in the action compared with standard Liouville action.}
\be
\la{4.7}
-\frac{3\pi}{2}\,\beta\,[1+\mc O(\beta^{2})] = c\,\kappa_{2}^{3}\,\mc C_{2}^{-3/2}\ \to \ c = \frac{12\pi}{\beta^{2}}
+\mc O(\beta^{0}).
\ee
As a further check, we move to the connected 4-point function. This is given by the diagrams in Fig.~(\ref{fig:A1}).
Their sum is 
\begin{align}
\la{4.8}
\mc C_{2}^{-2}\,\llangle & \varphi_{2}(t_{1})\cdots\varphi_{2}(t_{4})\rrangle_{\rm conn} = 16\,\beta^{2}\,(
W^{s}_{2222; 2}+W^{t}_{2222; 2}+W^{u}_{2222; 2} -\tfrac{1}{2}\,D_{2222})  \notag \\
&= 16\,\beta^{2}\,(
\tfrac{1}{4\,t_{12}^{2}}D_{1122}+\tfrac{1}{4\,t_{13}^{2}}D_{1212}+
\tfrac{1}{4\,t_{14}^{2}}D_{1221}-\tfrac{1}{2}\,D_{2222}) \notag \\
&= \tfrac{1}{t_{12}^{4}t_{34}^{4}}\tfrac{3\pi\beta^{2}}{2}\,\chi^{2}\,[\overline{D}_{1122}+\chi^{2}\,(
\overline{D}_{1212}+\overline{D}_{1221}-5\overline{D}_{2222})]
= \frac{1}{t_{12}^{4}t_{34}^{4}}\frac{3\,\pi\,\beta^{2}\,\chi^{2}\,(1-\chi+\chi^{2})}{2\,(1-\chi)^{2}}.
\end{align}
This can be written -- let us add explicitly the higher order corrections -- 
\be
\la{4.9}
\mc C_{2}^{-2}\,\llangle  \varphi_{2}(t_{1})\cdots\varphi_{2}(t_{4})\rrangle_{\rm conn} = \frac{3\pi\,\beta^{2}}{4}\,\bigg(
\frac{1}{t_{12}^{2}\,t_{23}^{2}\,t_{34}^{2}\,t_{14}^{2}}
+\frac{1}{t_{13}^{2}\,t_{24}^{2}\,t_{14}^{2}\,t_{23}^{2}}
+\frac{1}{t_{12}^{2}\,t_{24}^{2}\,t_{34}^{2}\,t_{13}^{2}}
\bigg)+\mc O(\beta^{4}),
\ee
and is consistent with the above identifications because the coefficient of the $\langle TTTT\rangle$ correlator is then
predicted to be 
\be
\kappa_{2}^{-4}\mc C_{2}^{2}\,\frac{3\pi\,\beta^{2}}{4}\,[1+\mc O(\beta^{2})] = c,
\ee
in agreement with (\ref{3.10}). \footnote{We choose $\kappa_{2}<0$ in order to have $\beta>0$.
Notice also that the disconnected diagrams give the $\mc O(c^{2})$ first term in (\ref{3.9}).}
Of course, the relation found in (\ref{4.7}) between the $c$ and $\beta$, {\em i.e.} 
\be
\la{4.12}
c = \frac{12\,\pi}{\beta^{2}}+\mc O(\beta^{0}),
\ee
should be considered as the leading order term at small $\beta$. The central charge of the $A_{n}$ Toda theory
is $c_{n}=n[1+(n+1)(n+2)(b+b^{-1})^{2}]$ where $b$ is proportional to $\beta$. With our conventions, {\em i.e.} requiring (\ref{4.12})
to hold at leading order for all $n$, this means that we could expect the following exact AdS/CFT map 
between the coupling $\beta$ and the central charge $c$
\be
c = n+12\,\pi\,\bigg[\frac{1}{\beta}+\frac{n(n+1)(n+2)}{12\,\pi}\,\beta\bigg]^{2} = 
\frac{12\pi}{\beta^{2}}+n\,(2n^{2}+6n+5)+\frac{n^{2}(n+1)^{2}(n+2)^{2}}{12\,\pi}\,\beta^{2}+\cdots.
\ee
As we reminded in the Introduction, the subleading $\mc O(\beta^{0})$ has been tested for the Liouville $A_{1}$
case in  \cite{Beccaria:2019stp}, see also \cite{BHT} for the $A_{2}$ theory and other generalizations complementary to 
this analysis.

\paragraph{Normalization of higher spin $s\ge 3$ duals}

If we assume that $\varphi_{s}=\kappa_{s}\,Q_{s}$ where $Q_{s}$ is a spin $s$ generator in a certain 
Virasoro extension $\mc W(2, \dots, s, \dots)$, then the same analysis of the 2-point function and the 
relation $\langle Q_{s}Q_{s}\rangle = \frac{c}{s}\frac{1}{t^{2s}}$, gives 
\be
\la{4.14}
\kappa_{s} = -\sqrt\frac{s}{c}+\mc O(\beta^{2}).
\ee
This implies a constraint on the vertices of the form $V=\beta\,g_{2ss}\,\varphi_{2}\varphi_{s}^{2}$. 
The associated Feynman rule is 
\begin{center}
\begin{tikzpicture}[line width=1 pt, scale=0.6]
\draw (0:1)--(0,0);       \node[right] at (0:1) {2};
\draw (120:1)--(0,0);   \node[left] at (120:1) {s};
\draw (-120:1)--(0,0);   \node[left] at (-120:1) {s};
\node[right=0.3cm] at (0:1) {$= -2\,\beta\,g_{2ss},$};
\end{tikzpicture}
\end{center}
then from (\ref{2.13})
\be
\mc C_{2}^{-1/2}\,\mc C_{s}^{-1}\,
\llangle \varphi_{2}(t_{1})\varphi_{s}(t_{2})\varphi_{s}(t_{3})\rrangle = (-2\beta g_{2ss})\times 
\frac{\sqrt\pi\,\Gamma(s+\frac{1}{2})}{2\,(s-1)\,\Gamma(s)}\,
\frac{1}{t_{12}^{2}\,t_{13}^{2}\,t_{23}^{2s-2}}\,(1+\mc O(\beta^{2})).
\ee
On the other hand,
\be
\mc C_{2}^{-1/2}\,\mc C_{s}^{-1}\,\llangle \varphi_{2}(t_{1})\varphi_{s}(t_{2})\varphi_{s}(t_{3})\rrangle  = 
\mc C_{2}^{-1/2}\,\mc C_{s}^{-1}\,
\kappa_{2}\kappa_{s}^{2}\,\langle T\,Q_{s}\,Q_{s}\rangle = \mc C_{2}^{-1/2}\,\mc C_{s}^{-1}\,
\frac{\kappa_{2}\kappa_{s}^{2}\,c}
{t_{12}^{2}\,t_{13}^{2}\,t_{23}^{2s-2}}.
\ee
From (\ref{4.5}) this gives for $s>2$ \footnote{The case $s=2$ is special because of the extra permutation
symmetry between the three $\varphi_{2}$ fields. In this case, the coefficient $\kappa$ is that in (\ref{4.5}).}
\be
\la{4.17}
\kappa_{s}^{2} =\frac{g_{2ss}}{c} \frac{2\,\sqrt\pi\,\Gamma(s+\frac{1}{2})}{(s-1)\,\Gamma(s)}\,\mc C_{s}
= \frac{g_{2ss}}{c}\,\frac{1}{s-1}\,(1+\mc O(\beta^{2})).
\ee
Using (\ref{4.14}), we find 
\be
g_{2ss} = s\,(s-1),
\ee
consistently with the explicit values in (\ref{2.8}),
$g_{233}=6$ and $g_{244}=12$.

\subsection{Matching the 4-point functions involving $\de=3,4$}

First of all, let us notice that the $\mc O(c^{2})$ terms in the CFT results (\ref{3.21}) immediately match 
the disconnected Witten diagrams where the four points on the boundary are connected with two boundary-to-boundary
propagators. The next correction is $\mc O(c)$ on the CFT side and should match the 
$\mc O(\beta^{2}/\kappa^{4})$ connected 4-point functions in AdS. \footnote{This is correct using (\ref{4.12}) and taking
into account the $\kappa^{4}\sim 1/c^{2}$ normalizations.}
A comparison of $G_{3333}^{\rm AdS}(\chi)$ with the CFT result (\ref{3.53}) valid for the $\mc W_{n}$
theory, shows that we have indeed 
\be
\frac{\beta^{2}\,G_{3333}^{\rm AdS}(\chi)}{c\,G_{3333,1}^{\mc W_{n+1}}(\chi)} = 
\frac{675\,\pi\,\beta^{2}}{256\,c}
=\mc C_{3}^{-2}\, (\kappa_{3})^{4},
\ee
where we used (\ref{4.12}) and (\ref{4.14}), and identified $z=\chi$. Similarly, 
comparing  $G_{4444}^{\rm AdS}(\chi)$ with the CFT result (\ref{3.57}) we find
\be
\frac{\beta^{2}\,G_{4444}^{\rm AdS}(\chi)}{c\,G_{4444, 1}^{\mc W_{n+1}}(\chi)} = 
\frac{1225\,\pi\,\beta^{2}}{192\,c}=\mc C_{4}^{-2}\, (\kappa_{4})^{4},
\ee
where we used (\ref{4.12}) and (\ref{4.14}), and identified $z=\chi$. Finally, for the mixed correlator,
we compare $G_{3344}^{\rm AdS}(\chi)$ with (\ref{3.55})  and have again 
\be
\frac{\beta^{2}\, G_{3344}^{\rm AdS}(\chi)}
{c\,G_{3344, 1}^{\mc W_{n+1}}(\chi)} = \frac{525\,\pi\,\beta^{2}}{128\,c}
= \mc C_{3}^{-1}\,\mc C_{4}^{-1}\,(\kappa_{3}\,\kappa_{4})^{2}.
\ee
These relations completes the proof that the four points functions of the $\Delta=3,4$ fields in the 
general $A_{n}$ Toda theory obey the relation (\ref{1.5}).

%
%

\section*{Acknowledgments}

We are very grateful to A. A. Tseytlin, S. Giombi, and H. Jiang
for many useful discussions related to the subject of this paper. 

\appendix

\section{Virasoro primary generating functions}

Let us briefly recall how Virasoro primaries are easily counted in $\mc W$-algebras by 
elementary character manipulations, see {\em e.g.} \cite{Candu:2012ne}. 
Acting on the vacuum with a bosonic primary with dimension $\de>0$, {\em i.e.} starting from 
$\phi_{-h}|0\rangle$, we obtain the associated Virasoro character
\be
\chi_{0,\de}=\prod_{n=0}^{\infty}\frac{1}{1-q^{\de+n}}.
\ee
The full character of a CFT can be decomposed in highest weight representations of Virasoro 
separating out the identity module and the contributions from primaries with dimensions $\de_{p}$
\be
\chi_{\text{full}} = \chi_{0,2}+\sum_{\de_{p}}d_{\de_{p}}\,q^{\de_{p}}\prod_{n=1}^{\infty}\frac{1}{1-q^{n}}.
\ee
From this relation one can extract the generating function of primary fields $\sum_{\de_{p}}d_{\de_{p}}\,q^{\de_{p}}$.
For example, to count the primaries in the $\mc W_{3}$ algebra we simply evaluate
\be
\la{A.3}
\sum_{\de_{p}}d_{\de_{p}} = \frac{\chi_{0,2}\chi_{0,3}-\chi_{0,2}}{\prod_{n=1}^{\infty}\frac{1}{1-q^{n}}}
= q^3+q^6+q^8+q^9+q^{10}+q^{11}+3 q^{12}+q^{13}+3 q^{14}+\cdots\,.
\ee
The first term is the primary $Q_{3}$, the second one is a composite $\sim (Q_{3}Q_{3})+\cdots$ . 
In the similar case of the $\mc W_{4}$ algebra we have 
\be
\la{A.4}
\sum_{\de_{p}}d_{\de_{p}} = \frac{\chi_{0,2}\chi_{0,3}\chi_{0,4}-\chi_{0,2}}{\prod_{n=1}^{\infty}\frac{1}{1-q^{n}}}
= q^3+q^4+q^6+q^7+3 q^8+2 q^9+4 q^{10}+\cdots.
\ee
The first terms are the spin 3 and 4 generators, the third is a composite $\sim (Q_{3}Q_{3})+\cdots$ . Of course, this is not 
the same as the dimension 6 primary of $\mc W_{3}$. In particular, it involves the spin 4 generator $Q_{4}$.

\section{Some AdS integrals}
\la{app-Dfun}

The basic $N$-point contact diagram connecting boundary points $\bm{t}=(t_{1}, \dots, t_{N})$  to the bulk point $(t,\bz)$
is given by 
\begin{align}
W_{\bm\de}(\bm t) = \int_{\text{AdS}_{2}}\frac{dt\,d\bz}{\bz^{2}}\,\prod_{i=1}^{N}
\left[\frac{\bz}{\bz^{2}+(t-t_{i})^{2}}\right]^{\de_{i}}.
\end{align}
Special cases are the 3-point function ($t_{ij}=t_{i}-t_{j}$)
\begin{align}
\la{2.13}
W_{\de_{1}\de_{2}\de_{3}}(t_{1},t_{2},t_{3}) = \frac{\sqrt\pi\,
\Gamma(\frac{\de_{1}+\de_{2}-\de_{3}}{2})\,
\Gamma(\frac{\de_{2}+\de_{3}-\de_{1}}{2})\,
\Gamma(\frac{\de_{1}+\de_{3}-\de_{2}}{2})\,
\Gamma(\frac{\de_{1}+\de_{2}+\de_{3}-1}{2})}
{2\,\Gamma(\de_{1})\,\Gamma(\de_{2})\,\Gamma(\de_{3})\ 
|t_{12}|^{\de_{1}+\de_{2}-\de_{3}}
|t_{13}|^{\de_{1}+\de_{3}-\de_{2}}
|t_{23}|^{\de_{2}+\de_{3}-\de_{1}}
},
\end{align}
and the 4-point function, {\em cf.} App. (\ref{app-Dfun}),
\begin{align}
\la{2.14}
W_{\de_{1}\de_{2}\de_{3}\de_{4}}(t_{1},t_{2},t_{3},t_{4}) \stackrel{\rm def}{=} 
D_{\de_{1}\de_{2}\de_{3}\de_{4}}(t_{1},t_{2},t_{3},t_{4}).
\end{align}
Here, $D$-functions are discussed in general in \cite{DHoker:1999kzh,Dolan:2000ut,Dolan:2003hv}. 
$D$-functions are related to the more convenient $\overline{D}$-functions that depend only on the (unique) 
conformally invariant cross-ratio. They are defined by 
\be
D_{\bm{\de}}(\bm{t}) = \frac{\sqrt\pi\,\Gamma(\Sigma-\frac{1}{2})}{2\,\prod_{n}\Gamma(\de_{n})}\,
\frac{t_{14}^{2\,(\Sigma-\de_{1}-\de_{4})}\,t_{34}^{2\,(\Sigma-\de_{3}-\de_{4})}}
{t_{13}^{2\,(\Sigma-\de_{4})}\,t_{24}^{2\,\de_{2}}}\,\overline{D}_{\bm{\de}}(\chi),
\ee
where $\Sigma = \frac{1}{2}\sum_{n}\de_{n}$ and \footnote{Any ambiguity associated with odd exponents 
should be resolved
by replacing $t_{ij}^{2X}\to (t_{ij}^{2})^{X}$ and considering that $\overline D$ is actually a function of $\chi^{2}$.}
\be
\chi = \frac{t_{12}\,t_{34}}{t_{13}\,t_{24}},\qquad t_{ij}=t_{i}-t_{j}.
\ee
For integer $\de_{n}$ we can evaluate all $\overline D$ functions by the recursion identities
of \cite{Arutyunov:2002fh}. 
In particular, for $0<\chi<1$, one has the special cases
\begin{align}
\la{B.4}
\overline{D}_{1133} &= -\tfrac{2 \chi ^2+3 \chi -3}{15 (1-\chi)^2}
-\tfrac{2 \chi ^4 \log\chi}{15 (1-\chi)^3}
-\tfrac{2 (\chi ^2+3 \chi +6)\,\log (1-\chi ) }{15 
\chi },\notag \\
\overline{D}_{2233} &= 
\tfrac{18 \chi ^4-29 \chi ^3+5 \chi ^2+48 \chi -24}{210 (1-\chi)^3 \chi ^2}
+\tfrac{(9 \chi ^2-28 \chi +28) \chi ^2 \log (\chi )}{105 (1-\chi)^4}
-\tfrac{(9 \chi ^3+8 \chi ^2+6 \chi +12) \log (1-\chi )}{105 \chi^3}, \notag \\
\overline{D}_{3333} &= 
-\tfrac{2 (12 \chi ^6-36 \chi ^5+37 \chi ^4-14 \chi ^3+37 \chi ^2-36 \chi +12)}{315 (1-\chi)^4 \chi ^4}
-\tfrac{4 (\chi ^2-\chi +1) (2 \chi ^2-7 \chi +7) \log (\chi )}{105 (1-\chi)^5}\notag \\
& -\tfrac{4 (\chi ^2-\chi +1) (2 \chi ^2+3 \chi +2) \log (1-\chi )}{105 \chi ^5},
\end{align}
with the additional cases related to the above by crossing
\begin{align}
\overline{D}_{1313} &= \tfrac{1}{(1-\chi)^{2}}\,\overline{D}_{1133}(\tfrac{1}{1-\chi}), & 
\overline{D}_{1331} &= \tfrac{1}{(1-\chi)^{4}}\,\overline{D}_{1133}(1-\chi), \notag  \\
\overline{D}_{2323} &= \tfrac{1}{(1-\chi)^{4}}\,\overline{D}_{2233}(\tfrac{1}{1-\chi}), &
\overline{D}_{2332} &= \tfrac{1}{(1-\chi)^{2}}\,\overline{D}_{2233}(\tfrac{1}{1-\chi}).
\end{align}

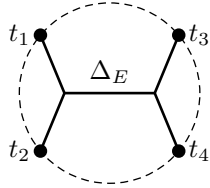
\begin{figure}[ht]
\centering
\begin{tikzpicture}[line width=1 pt, scale=0.6]
\coordinate (A1) at (140:2);
\coordinate (A2) at (-140:2);
\coordinate (A3) at (40:2);
\coordinate (A4) at (-40:2);
\coordinate (M1) at (-1,0);    \coordinate (M2) at (1,0);

\draw[densely dashed,line width=0.5pt] (0,0) circle (2);
\draw (A1)--(M1)--(A2); \draw (A3)--(M2)--(A4); \draw (M1)--(M2);

\draw[fill=black] (A1) circle (0.12); \node[left] at (A1) {$t_{1}$};
\draw[fill=black] (A2) circle (0.12); \node[left] at (A2) {$t_{2}$};
\draw[fill=black] (A3) circle (0.12); \node[right] at (A3) {$t_{3}$};
\draw[fill=black] (A4) circle (0.12); \node[right] at (A4) {$t_{4}$};


\node[above] at (0,0) {$\de_{E}$};
\end{tikzpicture}
\caption{Tree diagram associated with the s-channel exchange of a field with conformal parameter $\de_{E}$
mediated by two cubic vertices. The dashed line is just a convenient graphical representation for the compactified
boundary of AdS$_{2}$.
}\label{fig:exch}
\end{figure}

Besides the contact diagram expressed by  (\ref{2.14}), a generic 
boundary 4-point correlation function $\langle \Phi_{\de_{1}}(t_{1})\cdots \Phi_{\de_{4}}(t_{4})\rangle$
receives at tree level contributions from exchange diagrams mediated by two cubic
interactions, see Fig.~(\ref{fig:exch}).
In the s-channel and again with unit normalization, we have a simple formula valid when the exchanged 
field has conformal parameter $\de_{E}$ with $k=\frac{\de_{1}+\de_{2}-\de_{E}}{2}\in \mathbb N^{+}$.
In this case, the exchange diagram is given by \cite{DHoker:1999mqo}
\be
\la{2.15}
W^{s}_{\bm{\de}; \de_{E}}(\bm{t}) = \sum_{\ell=1}^{k}\frac{(\de_{1})_{-\ell}
(\de_{2})_{-\ell}}{4\,(k)_{1-\ell}(\frac{\de_{1}+\de_{2}-1+\de_{E}}{2})_{1-\ell}}\
\frac{1}{|t_{12}|^{2\ell}}\,D_{\de_{1}-\ell, \de_{2}-\ell, \de_{3}, \de_{4}}(\bm{t}),
\ee
with similar expressions for the other channels.

\bibliography{BT-Biblio}

\providecommand{\href}[2]{#2}\begingroup\raggedright\begin{thebibliography}{10}

\bibitem{Callan:1989em}
C.~G. Callan, Jr. and F.~Wilczek, \emph{{Iinfrared behaviour at negative
  curvature}},
  \href{http://dx.doi.org/10.1016/0550-3213(90)90451-I}{\emph{Nucl. Phys.} {\bf
  B340} (1990) 366--386}.

\bibitem{Paulos:2016fap}
M.~F. Paulos, J.~Penedones, J.~Toledo, B.~C. van Rees and P.~Vieira, \emph{{The
  S-matrix bootstrap. Part I: QFT in AdS}},
  \href{http://dx.doi.org/10.1007/JHEP11(2017)133}{\emph{JHEP} {\bf 11} (2017)
  133}, [\href{http://arxiv.org/abs/1607.06109}{{\tt 1607.06109}}].

\bibitem{Paulos:2016but}
M.~F. Paulos, J.~Penedones, J.~Toledo, B.~C. van Rees and P.~Vieira, \emph{{The
  S-matrix bootstrap II: two dimensional amplitudes}},
  \href{http://dx.doi.org/10.1007/JHEP11(2017)143}{\emph{JHEP} {\bf 11} (2017)
  143}, [\href{http://arxiv.org/abs/1607.06110}{{\tt 1607.06110}}].

\bibitem{Carmi:2018qzm}
D.~Carmi, L.~Di~Pietro and S.~Komatsu, \emph{{A Study of Quantum Field Theories
  in AdS at Finite Coupling}},  \href{http://arxiv.org/abs/1810.04185}{{\tt
  1810.04185}}.

\bibitem{DHoker:1983zwg}
E.~D'Hoker and R.~Jackiw, \emph{{Space translation breaking and
  compactification in the Liouville theory}},
  \href{http://dx.doi.org/10.1103/PhysRevLett.50.1719}{\emph{Phys. Rev. Lett.}
  {\bf 50} (1983) 1719--1722}.

\bibitem{DHoker:1983msr}
E.~D'Hoker, D.~Z. Freedman and R.~Jackiw, \emph{{$SO(2,1)$ Invariant
  Quantization of the Liouville Theory}},
  \href{http://dx.doi.org/10.1103/PhysRevD.28.2583}{\emph{Phys. Rev.} {\bf D28}
  (1983) 2583}.

\bibitem{Inami:1985di}
T.~Inami and H.~Ooguri, \emph{{Dynamical breakdown of sypersymmetry in
  two-dimensional Anti de Sitter space }},
  \href{http://dx.doi.org/10.1016/0550-3213(86)90255-5}{\emph{Nucl. Phys.} {\bf
  B273} (1986) 487--500}.

\bibitem{Strominger:1998yg}
A.~Strominger, \emph{{AdS$_{2}$ quantum gravity and string theory}},
  \href{http://dx.doi.org/10.1088/1126-6708/1999/01/007}{\emph{JHEP} {\bf 01}
  (1999) 007}, [\href{http://arxiv.org/abs/hep-th/9809027}{{\tt
  hep-th/9809027}}].

\bibitem{Drukker:2000ep}
N.~Drukker, D.~J. Gross and A.~A. Tseytlin, \emph{{Green-Schwarz string in
  $\text{AdS}_{5}\times S^{5}$: Semiclassical partition function}},
  \href{http://dx.doi.org/10.1088/1126-6708/2000/04/021}{\emph{JHEP} {\bf 04}
  (2000) 021}, [\href{http://arxiv.org/abs/hep-th/0001204}{{\tt
  hep-th/0001204}}].

\bibitem{Alday:2007he}
L.~F. Alday and J.~Maldacena, \emph{{Comments on gluon scattering amplitudes
  via AdS/CFT}},
  \href{http://dx.doi.org/10.1088/1126-6708/2007/11/068}{\emph{JHEP} {\bf 11}
  (2007) 068}, [\href{http://arxiv.org/abs/0710.1060}{{\tt 0710.1060}}].

\bibitem{Polyakov:2000ti}
A.~M. Polyakov and V.~S. Rychkov, \emph{{Gauge field strings duality and the
  loop equation}},
  \href{http://dx.doi.org/10.1016/S0550-3213(00)00183-8}{\emph{Nucl. Phys.}
  {\bf B581} (2000) 116--134}, [\href{http://arxiv.org/abs/hep-th/0002106}{{\tt
  hep-th/0002106}}].

\bibitem{Polchinski:2011im}
J.~Polchinski and J.~Sully, \emph{{Wilson Loop Renormalization Group Flows}},
  \href{http://dx.doi.org/10.1007/JHEP10(2011)059}{\emph{JHEP} {\bf 10} (2011)
  059}, [\href{http://arxiv.org/abs/1104.5077}{{\tt 1104.5077}}].

\bibitem{Drukker:2006xg}
N.~Drukker and S.~Kawamoto, \emph{{Small deformations of supersymmetric Wilson
  loops and open spin-chains}},
  \href{http://dx.doi.org/10.1088/1126-6708/2006/07/024}{\emph{JHEP} {\bf 07}
  (2006) 024}, [\href{http://arxiv.org/abs/hep-th/0604124}{{\tt
  hep-th/0604124}}].

\bibitem{Giombi:2017cqn}
S.~Giombi, R.~Roiban and A.~A. Tseytlin, \emph{{Half-BPS Wilson loop and
  AdS$_2$/CFT$_1$}},
  \href{http://dx.doi.org/10.1016/j.nuclphysb.2017.07.004}{\emph{Nucl. Phys.}
  {\bf B922} (2017) 499--527}, [\href{http://arxiv.org/abs/1706.00756}{{\tt
  1706.00756}}].

\bibitem{Beccaria:2018ocq}
M.~Beccaria and A.~A. Tseytlin, \emph{{On non-supersymmetric generalizations of
  the Wilson-Maldacena loops in $\mathcal{N}=4$ SYM}},
  \href{http://dx.doi.org/10.1016/j.nuclphysb.2018.07.019}{\emph{Nucl. Phys.}
  {\bf B934} (2018) 466--497}, [\href{http://arxiv.org/abs/1804.02179}{{\tt
  1804.02179}}].

\bibitem{Beccaria:2019dws}
M.~Beccaria, S.~Giombi and A.~A. Tseytlin, \emph{{Correlators on
  non-supersymmetric Wilson line in $\mathcal N=4$ SYM and AdS$_2$/CFT$_1$}},
  \href{http://arxiv.org/abs/1903.04365}{{\tt 1903.04365}}.

\bibitem{Hotta:1998iq}
M.~Hotta, \emph{{Asymptotic isometry and two-dimensional anti-de Sitter
  gravity}},  \href{http://arxiv.org/abs/gr-qc/9809035}{{\tt gr-qc/9809035}}.

\bibitem{Cadoni:1999ja}
M.~Cadoni and S.~Mignemi, \emph{{Asymptotic symmetries of AdS$_{2}$ and
  conformal group in d = 1}},
  \href{http://dx.doi.org/10.1016/S0550-3213(99)00398-3}{\emph{Nucl. Phys.}
  {\bf B557} (1999) 165--180}, [\href{http://arxiv.org/abs/hep-th/9902040}{{\tt
  hep-th/9902040}}].

\bibitem{NavarroSalas:1999up}
J.~Navarro-Salas and P.~Navarro, \emph{{AdS$_{2}$/CFT$_{1}$ correspondence and
  near extremal black hole entropy}},
  \href{http://dx.doi.org/10.1016/S0550-3213(00)00165-6}{\emph{Nucl. Phys.}
  {\bf B579} (2000) 250--266}, [\href{http://arxiv.org/abs/hep-th/9910076}{{\tt
  hep-th/9910076}}].

\bibitem{Almheiri:2014cka}
A.~Almheiri and J.~Polchinski, \emph{{Models of AdS$_{2}$ backreaction and
  holography}}, \href{http://dx.doi.org/10.1007/JHEP11(2015)014}{\emph{JHEP}
  {\bf 11} (2015) 014}, [\href{http://arxiv.org/abs/1402.6334}{{\tt
  1402.6334}}].

\bibitem{Maldacena:2016upp}
J.~Maldacena, D.~Stanford and Z.~Yang, \emph{{Conformal symmetry and its
  breaking in two dimensional Nearly Anti-de-Sitter space}},
  \href{http://dx.doi.org/10.1093/ptep/ptw124}{\emph{PTEP} {\bf 2016} (2016)
  12C104}, [\href{http://arxiv.org/abs/1606.01857}{{\tt 1606.01857}}].

\bibitem{Engelsoy:2016xyb}
J.~Engelsoy, T.~G. Mertens and H.~Verlinde, \emph{{An investigation of
  AdS$_{2}$ backreaction and holography}},
  \href{http://dx.doi.org/10.1007/JHEP07(2016)139}{\emph{JHEP} {\bf 07} (2016)
  139}, [\href{http://arxiv.org/abs/1606.03438}{{\tt 1606.03438}}].

\bibitem{Beccaria:2019stp}
M.~Beccaria and A.~A. Tseytlin, \emph{{On boundary correlators in Liouville
  theory on AdS$_{2}$}},  \href{http://arxiv.org/abs/1904.12753}{{\tt
  1904.12753}}.

\bibitem{Polyakov:1981rd}
A.~M. Polyakov, \emph{{Quantum Geometry of Bosonic Strings}},
  \href{http://dx.doi.org/10.1016/0370-2693(81)90743-7}{\emph{Phys.Lett.} {\bf
  B103} (1981) 207--210}.

\bibitem{Teschner:2001rv}
J.~Teschner, \emph{{Liouville theory revisited}},
  \href{http://dx.doi.org/10.1088/0264-9381/18/23/201}{\emph{Class. Quant.
  Grav.} {\bf 18} (2001) R153--R222},
  [\href{http://arxiv.org/abs/hep-th/0104158}{{\tt hep-th/0104158}}].

\bibitem{Nakayama:2004vk}
Y.~Nakayama, \emph{{Liouville field theory: A Decade after the revolution}},
  \href{http://dx.doi.org/10.1142/S0217751X04019500}{\emph{Int. J. Mod. Phys.}
  {\bf A19} (2004) 2771--2930},
  [\href{http://arxiv.org/abs/hep-th/0402009}{{\tt hep-th/0402009}}].

\bibitem{Zamolodchikov:2001ah}
A.~B. Zamolodchikov and A.~B. Zamolodchikov, \emph{{Liouville field theory on a
  pseudosphere}},  \href{http://arxiv.org/abs/hep-th/0101152}{{\tt
  hep-th/0101152}}.

\bibitem{Menotti:2004uq}
P.~Menotti and E.~Tonni, \emph{{Standard and geometric approaches to quantum
  Liouville theory on the pseudosphere}},
  \href{http://dx.doi.org/10.1016/j.nuclphysb.2004.11.003}{\emph{Nucl. Phys.}
  {\bf B707} (2005) 321--346}, [\href{http://arxiv.org/abs/hep-th/0406014}{{\tt
  hep-th/0406014}}].

\bibitem{Ouyang:2019xdd}
H.~Ouyang, \emph{{Holographic four-point functions in Toda field theories in
  AdS$_{2}$}}, \href{http://dx.doi.org/10.1007/JHEP04(2019)159}{\emph{JHEP}
  {\bf 04} (2019) 159}, [\href{http://arxiv.org/abs/1902.10536}{{\tt
  1902.10536}}].

\bibitem{Gervais:1983am}
J.-L. Gervais and A.~Neveu, \emph{{New Quantum Treatment of Liouville Field
  Theory}}, \href{http://dx.doi.org/10.1016/0550-3213(83)90008-1}{\emph{Nucl.
  Phys.} {\bf B224} (1983) 329--348}.

\bibitem{Mansfield:1982sq}
P.~Mansfield, \emph{{Light Cone Quantization of the Liouville and Toda Field
  Theories}}, \href{http://dx.doi.org/10.1016/0550-3213(83)90543-6}{\emph{Nucl.
  Phys.} {\bf B222} (1983) 419--445}.

\bibitem{Braaten:1983pz}
E.~Braaten, T.~Curtright, G.~Ghandour and C.~B. Thorn, \emph{{A Class of
  Conformally Invariant Quantum Field Theories}},
  \href{http://dx.doi.org/10.1016/0370-2693(83)91288-1}{\emph{Phys. Lett.} {\bf
  125B} (1983) 301--304}.

\bibitem{BHT}
M.~Beccaria, H.~Jiang and A.~A. Tseytlin, \emph{{in preparation}}, .

\bibitem{Bouwknegt:1992wg}
P.~Bouwknegt and K.~Schoutens, \emph{{W symmetry in conformal field theory}},
  \href{http://dx.doi.org/10.1016/0370-1573(93)90111-P}{\emph{Phys. Rept.} {\bf
  223} (1993) 183--276}, [\href{http://arxiv.org/abs/hep-th/9210010}{{\tt
  hep-th/9210010}}].

\bibitem{Fateev:1987zh}
V.~A. Fateev and S.~L. Lukyanov, \emph{{The Models of Two-Dimensional Conformal
  Quantum Field Theory with Z$_{n}$ Symmetry}},
  \href{http://dx.doi.org/10.1142/S0217751X88000205}{\emph{Int. J. Mod. Phys.}
  {\bf A3} (1988) 507}.

\bibitem{Fateev:2007ab}
V.~A. Fateev and A.~V. Litvinov, \emph{{Correlation functions in conformal Toda
  field theory. I.}},
  \href{http://dx.doi.org/10.1088/1126-6708/2007/11/002}{\emph{JHEP} {\bf 11}
  (2007) 002}, [\href{http://arxiv.org/abs/0709.3806}{{\tt 0709.3806}}].

\bibitem{Christe:1989ah}
P.~Christe and G.~Mussardo, \emph{{Integrable Systems Away from Criticality:
  The Toda Field Theory and S Matrix of the Tricritical Ising Model}},
  \href{http://dx.doi.org/10.1016/0550-3213(90)90119-X}{\emph{Nucl. Phys.} {\bf
  B330} (1990) 465--487}.

\bibitem{Christe:1989my}
P.~Christe and G.~Mussardo, \emph{{Elastic s Matrices in $(1+1)$-Dimensions and
  Toda Field Theories}},
  \href{http://dx.doi.org/10.1142/S0217751X90001938}{\emph{Int. J. Mod. Phys.}
  {\bf A5} (1990) 4581--4628}.

\bibitem{Braden:1989bu}
H.~W. Braden, E.~Corrigan, P.~E. Dorey and R.~Sasaki, \emph{{Affine Toda Field
  Theory and Exact S Matrices}},
  \href{http://dx.doi.org/10.1016/0550-3213(90)90648-W}{\emph{Nucl. Phys.} {\bf
  B338} (1990) 689--746}.

\bibitem{Braden:1989bg}
H.~W. Braden, E.~Corrigan, P.~E. Dorey and R.~Sasaki, \emph{{Extended Toda
  Field Theory and Exact S Matrices}},
  \href{http://dx.doi.org/10.1016/0370-2693(89)90952-0}{\emph{Phys. Lett.} {\bf
  B227} (1989) 411--416}.

\bibitem{Gabai:2018tmm}
B.~Gabai, D.~Mazac, A.~Shieber, P.~Vieira and Y.~Zhou, \emph{{No Particle
  Production in Two Dimensions: Recursion Relations and Multi-Regge Limit}},
  \href{http://dx.doi.org/10.1007/JHEP02(2019)094}{\emph{JHEP} {\bf 02} (2019)
  094}, [\href{http://arxiv.org/abs/1803.03578}{{\tt 1803.03578}}].

\bibitem{Dolan:2000ut}
F.~A. Dolan and H.~Osborn, \emph{{Conformal four point functions and the
  operator product expansion}},
  \href{http://dx.doi.org/10.1016/S0550-3213(01)00013-X}{\emph{Nucl. Phys.}
  {\bf B599} (2001) 459--496}, [\href{http://arxiv.org/abs/hep-th/0011040}{{\tt
  hep-th/0011040}}].

\bibitem{Dolan:2003hv}
F.~A. Dolan and H.~Osborn, \emph{{Conformal partial waves and the operator
  product expansion}},
  \href{http://dx.doi.org/10.1016/j.nuclphysb.2003.11.016}{\emph{Nucl. Phys.}
  {\bf B678} (2004) 491--507}, [\href{http://arxiv.org/abs/hep-th/0309180}{{\tt
  hep-th/0309180}}].

\bibitem{Perlmutter:2015iya}
E.~Perlmutter, \emph{{Virasoro conformal blocks in closed form}},
  \href{http://dx.doi.org/10.1007/JHEP08(2015)088}{\emph{JHEP} {\bf 08} (2015)
  088}, [\href{http://arxiv.org/abs/1502.07742}{{\tt 1502.07742}}].

\bibitem{Zamolodchikov:1985wn}
A.~B. Zamolodchikov, \emph{{Infinite Additional Symmetries in Two-Dimensional
  Conformal Quantum Field Theory}},
  \href{http://dx.doi.org/10.1007/BF01036128}{\emph{Theor. Math. Phys.} {\bf
  65} (1985) 1205--1213}.

\bibitem{Bowcock:1991zk}
P.~Bowcock and G.~M.~T. Watts, \emph{{On the classification of quantum
  $\mathcal W$ algebras}},
  \href{http://dx.doi.org/10.1016/0550-3213(92)90590-8}{\emph{Nucl. Phys.} {\bf
  B379} (1992) 63--95}, [\href{http://arxiv.org/abs/hep-th/9111062}{{\tt
  hep-th/9111062}}].

\bibitem{Fateev:1987vh}
V.~A. Fateev and A.~B. Zamolodchikov, \emph{{Conformal Quantum Field Theory
  Models in Two-Dimensions Having Z$_{3}$ Symmetry}},
  \href{http://dx.doi.org/10.1016/0550-3213(87)90166-0}{\emph{Nucl. Phys.} {\bf
  B280} (1987) 644--660}.

\bibitem{Bais:1987dc}
F.~Bais, P.~Bouwknegt, M.~Surridge and K.~Schoutens, \emph{{Extensions of the
  Virasoro Algebra Constructed from Kac-Moody Algebras Using Higher Order
  Casimir Invariants}},
  \href{http://dx.doi.org/10.1016/0550-3213(88)90631-1}{\emph{Nucl.Phys.} {\bf
  B304} (1988) 348--370}.

\bibitem{Bais:1987zk}
F.~A. Bais, P.~Bouwknegt, M.~Surridge and K.~Schoutens, \emph{{Coset
  Construction for Extended Virasoro Algebras}},
  \href{http://dx.doi.org/10.1016/0550-3213(88)90632-3}{\emph{Nucl. Phys.} {\bf
  B304} (1988) 371--391}.

\bibitem{Kausch:1990bn}
H.~Kausch and G.~Watts, \emph{{A Study of W algebras using Jacobi identities}},
  \href{http://dx.doi.org/10.1016/0550-3213(91)90375-8}{\emph{Nucl.Phys.} {\bf
  B354} (1991) 740--768}.

\bibitem{Thielemans:1991uw}
K.~Thielemans, \emph{{A Mathematica package for computing operator product
  expansions}}, \href{http://dx.doi.org/10.1142/S0129183191001001}{\emph{Int.
  J. Mod. Phys.} {\bf C2} (1991) 787--798}.

\bibitem{Fitzpatrick:2016mtp}
A.~L. Fitzpatrick, J.~Kaplan, D.~Li and J.~Wang, \emph{{Exact Virasoro Blocks
  from Wilson Lines and Background-Independent Operators}},
  \href{http://dx.doi.org/10.1007/JHEP07(2017)092}{\emph{JHEP} {\bf 07} (2017)
  092}, [\href{http://arxiv.org/abs/1612.06385}{{\tt 1612.06385}}].

\bibitem{Hikida:2017ehf}
Y.~Hikida and T.~Uetoko, \emph{{Correlators in higher-spin $AdS_3$ holography
  from Wilson lines with loop corrections}},
  \href{http://dx.doi.org/10.1093/ptep/ptx154}{\emph{PTEP} {\bf 2017} (2017)
  113B03}, [\href{http://arxiv.org/abs/1708.08657}{{\tt 1708.08657}}].

\bibitem{Hikida:2018dxe}
Y.~Hikida and T.~Uetoko, \emph{{Conformal blocks from Wilson lines with loop
  corrections}},
  \href{http://dx.doi.org/10.1103/PhysRevD.97.086014}{\emph{Phys. Rev.} {\bf
  D97} (2018) 086014}, [\href{http://arxiv.org/abs/1801.08549}{{\tt
  1801.08549}}].

\bibitem{Bombini:2018jrg}
A.~Bombini, S.~Giusto and R.~Russo, \emph{{A note on the Virasoro blocks at
  order $1/c$}},
  \href{http://dx.doi.org/10.1140/epjc/s10052-018-6522-5}{\emph{Eur. Phys. J.}
  {\bf C79} (2019) 3}, [\href{http://arxiv.org/abs/1807.07886}{{\tt
  1807.07886}}].

\bibitem{Hornfeck:1992he}
K.~Hornfeck, \emph{{The Minimal supersymmetric extension of WA(n-1)}},
  \href{http://dx.doi.org/10.1016/0370-2693(92)91602-6}{\emph{Phys. Lett.} {\bf
  B275} (1992) 355--360}.

\bibitem{Hornfeck:1993kp}
K.~Hornfeck, \emph{{Classification of structure constants for W algebras from
  highest weights}},
  \href{http://dx.doi.org/10.1016/0550-3213(94)90061-2}{\emph{Nucl. Phys.} {\bf
  B411} (1994) 307--320}, [\href{http://arxiv.org/abs/hep-th/9307170}{{\tt
  hep-th/9307170}}].

\bibitem{Blumenhagen:1994wg}
R.~Blumenhagen, W.~Eholzer, A.~Honecker, K.~Hornfeck and R.~Hubel, \emph{{Coset
  realization of unifying W algebras}},
  \href{http://dx.doi.org/10.1142/S0217751X95001157}{\emph{Int. J. Mod. Phys.}
  {\bf A10} (1995) 2367--2430},
  [\href{http://arxiv.org/abs/hep-th/9406203}{{\tt hep-th/9406203}}].

\bibitem{Linshaw:2017tvv}
A.~R. Linshaw, \emph{{Universal two-parameter $\mathcal{W}_{\infty}$-algebra
  and vertex algebras of type $\mathcal{W}(2,3,\dots, N)$}},
  \href{http://arxiv.org/abs/1710.02275}{{\tt 1710.02275}}.

\bibitem{Gaberdiel:2012ku}
M.~R. Gaberdiel and R.~Gopakumar, \emph{{Triality in Minimal Model
  Holography}}, \href{http://dx.doi.org/10.1007/JHEP07(2012)127}{\emph{JHEP}
  {\bf 1207} (2012) 127}, [\href{http://arxiv.org/abs/1205.2472}{{\tt
  1205.2472}}].

\bibitem{Candu:2012ne}
C.~Candu, M.~R. Gaberdiel, M.~Kelm and C.~Vollenweider, \emph{{Even spin
  minimal model holography}},
  \href{http://dx.doi.org/10.1007/JHEP01(2013)185}{\emph{JHEP} {\bf 01} (2013)
  185}, [\href{http://arxiv.org/abs/1211.3113}{{\tt 1211.3113}}].

\bibitem{DHoker:1999kzh}
E.~D'Hoker, D.~Z. Freedman, S.~D. Mathur, A.~Matusis and L.~Rastelli,
  \emph{{Graviton exchange and complete four point functions in the AdS/CFT
  correspondence}},
  \href{http://dx.doi.org/10.1016/S0550-3213(99)00525-8}{\emph{Nucl. Phys.}
  {\bf B562} (1999) 353--394}, [\href{http://arxiv.org/abs/hep-th/9903196}{{\tt
  hep-th/9903196}}].

\bibitem{Arutyunov:2002fh}
G.~Arutyunov, F.~A. Dolan, H.~Osborn and E.~Sokatchev, \emph{{Correlation
  functions and massive Kaluza-Klein modes in the AdS/CFT correspondence}},
  \href{http://dx.doi.org/10.1016/S0550-3213(03)00448-6}{\emph{Nucl. Phys.}
  {\bf B665} (2003) 273--324}, [\href{http://arxiv.org/abs/hep-th/0212116}{{\tt
  hep-th/0212116}}].

\bibitem{DHoker:1999mqo}
E.~D'Hoker, D.~Z. Freedman and L.~Rastelli, \emph{{AdS/CFT four point
  functions: How to succeed at z integrals without really trying}},
  \href{http://dx.doi.org/10.1016/S0550-3213(99)00526-X}{\emph{Nucl. Phys.}
  {\bf B562} (1999) 395--411}, [\href{http://arxiv.org/abs/hep-th/9905049}{{\tt
  hep-th/9905049}}].

\end{thebibliography}\endgroup
\bibliographystyle{JHEP}

\end{document}